\documentclass{article}
\usepackage{epsfig}

\date{\today}

\usepackage{amsmath}
\usepackage{amssymb}
\usepackage[hang,nooneline,scriptsize]{subfigure}
\renewcommand{\thefootnote}{\arabic{footnote}}

\makeatletter
\let\@fnsymbol\@arabic
\makeatother

\begin{document}
\title{Self-interacting boson stars with a single Killing vector field in Anti-de Sitter}

\author{{\large Yves Brihaye 
}$^{(1)}$, {\large Betti Hartmann}$^{(2)}$
{\large J\"urgen Riedel }$^{(3)}$\\
\\ 
$^{(1)}${\small Physique-Math\'ematique, Universit\'e de
Mons, 7000 Mons, Belgium}\\ 
$^{(2)}${\small Instituto de F\'isica de S\~ao Carlos, Universidade de S\~ao Paulo, Caixa Postal 369, CEP 13560-970, S\~ao Carlos (SP), Brazil}\\ 
$^{(3)}${ \small LABORES - Laboratoire de Recherche Scientifique, 75006 Paris, France} }

\date{\today}
\setlength{\footnotesep}{0.5\footnotesep}
\newcommand{\dd}{\mbox{d}}
\newcommand{\tr}{\mbox{tr}}
\newcommand{\la}{\lambda}
\newcommand{\ka}{\kappa}
\newcommand{\f}{\phi}
\newcommand{\vf}{\varphi}
\newcommand{\F}{\Phi}
\newcommand{\al}{\alpha}
\newcommand{\ga}{\gamma}
\newcommand{\de}{\delta}
\newcommand{\si}{\sigma}
\newcommand{\bomega}{\mbox{\boldmath $\omega$}}
\newcommand{\bsi}{\mbox{\boldmath $\sigma$}}
\newcommand{\bchi}{\mbox{\boldmath $\chi$}}
\newcommand{\bal}{\mbox{\boldmath $\alpha$}}
\newcommand{\bpsi}{\mbox{\boldmath $\psi$}}
\newcommand{\brho}{\mbox{\boldmath $\varrho$}}
\newcommand{\beps}{\mbox{\boldmath $\varepsilon$}}
\newcommand{\bxi}{\mbox{\boldmath $\xi$}}
\newcommand{\bbeta}{\mbox{\boldmath $\beta$}}
\newcommand{\ee}{\end{equation}}
\newcommand{\eea}{\end{eqnarray}}
\newcommand{\be}{\begin{equation}}
\newcommand{\bea}{\begin{eqnarray}}

\newcommand{\ii}{\mbox{i}}
\newcommand{\e}{\mbox{e}}
\newcommand{\pa}{\partial}
\newcommand{\Om}{\Omega}
\newcommand{\vep}{\varepsilon}
\newcommand{\bfph}{{\bf \phi}}
\newcommand{\lm}{\lambda}
\def\theequation{\arabic{equation}}
\renewcommand{\thefootnote}{\fnsymbol{footnote}}
\newcommand{\re}[1]{(\ref{#1})}
\newcommand{\R}{{\rm I \hspace{-0.52ex} R}}
\newcommand{\N}{{\sf N\hspace*{-1.0ex}\rule{0.15ex}%
{1.3ex}\hspace*{1.0ex}}}
\newcommand{\Q}{{\sf Q\hspace*{-1.1ex}\rule{0.15ex}%
{1.5ex}\hspace*{1.1ex}}}
\newcommand{\C}{{\sf C\hspace*{-0.9ex}\rule{0.15ex}%
{1.3ex}\hspace*{0.9ex}}}
\newcommand{\eins}{1\hspace{-0.56ex}{\rm I}}
\renewcommand{\thefootnote}{\arabic{footnote}}
 \maketitle
\begin{abstract} 
We construct rotating boson stars in (4+1)-dimensional asymptotically Anti-de Sitter space-time (aAdS) with two
equal angular momenta that are composed out of a massive and self-interacting scalar field.
These solutions possess a single Killing vector field.
We construct explicit solutions of the equations in the case of a fixed AdS background and vanishing
self-coupling of the scalar field. These are the generalizations of the oscillons discussed in the literature
previously now taking the mass of the scalar field into account. We study the evolution of the spectrum
of massive oscillons when taking backreaction and/or the self-coupling into account numerically.
We observe that very compact boson stars possess an ergoregion.

\end{abstract}

\section{Introduction}
Anti-de Sitter space-time (AdS) is the unique ground state in the family of all possible asymptotically AdS space-times (aAdS).
It was shown that Minkowski space-time is non-linearly stable \cite{Christodoulou_Klainerman} and hence perturbations
stay small due to dissipation. While 
AdS is linearly stable \cite{Ishibashi:2004wx} a general result on the non-linear stability does not yet
exist for AdS. In fact, since AdS can be seen as a finite box, one would expect small perturbations to bounce of the
conformal boundary, to interact with themselves and lead to instabilities. This problem was recently tackled numerically
and it was conjectured that a large class of arbitrarily small scalar perturbations in $(3+1)$-dimensional AdS leads to the formation
of black holes \cite{Bizon:2011gg}. This was later extended to higher dimensions \cite{Jalmuzna:2011qw} 
It was found that this is related to the fact that energy is transferred to smaller and smaller
scales and was confirmed for gravitational perturbations as well \cite{Dias:2011ss}. This is a quite interesting observation since the
formation of a black hole in AdS can be interpreted as thermalization in the boundary field theory via the AdS/CFT correspondence 
\cite{ads_cft}. Despite this evidence a large number of aAdS solutions exists that are non-linearly stable \cite{Dias:2012tq}.
The prime example are boson stars in aAdS. Boson stars are solutions to gravitating scalar field models and exist due to their
periodic time-dependence which leads to a conserved Noether charge related to the $U(1)$ symmetry of the model. 
Boson stars can be constructed with and without scalar potential and boson stars in aAdS exist even for massless scalar fields
without self-interaction. To distinguish these from the solutions that we discuss in our paper we refer to these latter solutions in the following as
{\it minimal boson stars}. For minimal boson stars Newton's constant as well as the AdS radius can be scaled to unity and no (dimensionless)
coupling constants exist that can be varied continuously. This is different for a massive scalar field with self-interaction studied in this paper.
The potential parameters can be scaled to unity such that two dimensionless couplings remain: the ratio between the energy scale of the scalar
field and the Planck mass as well as the ratio between the mass of the scalar field and the negative cosmological constant. Then,
families of solutions can be constructed that display a much richer spectrum than for the minimal case. 

Spherically symmetric, non-rotating minimal boson stars in aAdS have been first discussed in 
\cite{Astefanesei:2003qy} and for a massive, self-interacting scalar field and in a various number
of spatial dimensions in \cite{Hartmann:2012wa,Hartmann:2012gw}.
The non-linear stability of spherically symmetric, non-rotating minimal boson stars in (3+1)-dimensional aAdS has been discussed 
in \cite{Buchel:2013uba,Maliborski:2013ula} and it was found that
the fundamental as well as the first few radially excited solutions are non-linearly stable. 

Most studies of boson stars in aAdS have been concerned with non-rotating solutions.
However, studies of rotating and/or non-spherically
symmetric solutions in aAdS might shed further light on the stability of AdS. As such rotating boson stars in (2+1)-dimensional
aAdS have been discussed \cite{Astefanesei:2003rw} as well as axially symmetric, rotating charged boson stars in
(3+1)-dimensional aAdS \cite{Kichakova:2013sza}. 
In (4+1)-dimensional asymptotically flat space-time rotating boson stars possessing only
one Killing field have been constructed \cite{Hartmann:2010pm}. 
The model consists of a doublet of complex scalar fields minimally coupled
to gravity which is only invariant under a specific linear combination of the Killing symmetries of the metric. 
This idea was used to construct black holes with scalar hair in (4+1)-dimensional aAdS
that are neither axisymmetric nor stationary \cite{Dias:2011at}. It was also shown 
that Kerr black holes with scalar hair in 4-dimensional asymptotically flat space-time
can be constructed using similar principles \cite{Herdeiro:2014goa}. This demonstrates that the existence of black holes
that are neither axisymmetric nor stationary is not a peculiarity of aAdS.

Rotating minimal boson stars in aAdS have been constructed in higher dimensions
\cite{Stotyn:2013yka}. It is also of interest (with view to the holographic interpretation in the context of the AdS/CFT correspondence) 
to construct aAdS solutions in higher order gravity models. The first step in this direction was taken in \cite{hartmann_riedel_suciu} with the 
construction of Gauss-Bonnet (GB) boson stars in asymptotically flat space-time. This was extended to rotating GB boson stars including
the aAdS case \cite{Brihaye:2013zha}. In this latter paper it was
conjectured that rotating GB boson stars do not exist for large values of the GB coupling.
 This conjecture has been confirmed recently for $d=5,6,7,11$ space-time dimensions
 in \cite{Henderson:2014dwa} for minimal GB boson stars. A perturbative 
 as well as numerical approach has been used. The expansion of the solutions
 indeed demonstrates that the solutions 
 cease to exist at a critical value of the GB coupling.

The aim of this paper is to discuss the non-perturbative aAdS rotating boson star solutions of a model with a massive, self-interacting scalar field.
Next to the fundamental solutions we also construct radially excited solutions. 

Solutions that possess angular momentum can suffer from 
a superradiant instability \cite{Press:1972zz}. If 
an ergoregion is present in the space-time of these rotating solutions infalling
bosonic waves are amplified when reflected. This is very similar to the Penrose process which
allows for energy extraction from a rotating black hole since the asymptotically time-like Killing
vector becomes space-like inside the ergoregion. As pointed out in \cite{Press:1972zz} the important ingredient is the presence of a
reflecting mirror. Since aAdS can act as such a mirror one would expect such instabilities to be present 
for rotating black holes in AdS. As such the Kerr-AdS black hole was conjectured to be unstable
for large rotation parameter $a=J/M$ \cite{Hawking:1999dp} and the issue was further discussed in 
\cite{Cardoso:2006wa,Cardoso:2004hs,Cardoso:2004nk}.
While here we are not dealing with black holes, we would expect that the existence of ergoregions
leads to a similar amplification of scalar bosonic waves as in the case of black holes.
We have hence investigated whether such ergoregions exist in our case, however leave the discussion of the superradiant instabilitiy
of our solutions for future work.

Our paper is organised as follows: in Section 2, we discuss the model. In Section 3 we discuss the solutions
and put specific emphasis on the stability of the boson stars.
In Section 4 we conclude.

\section{The model}
The model we are using for the construction of rotating and radially excited boson stars in aAdS reads
\begin{equation}
\label{action}
S= \frac{1}{16\pi G} \int d^5 x \sqrt{-g_m} \left(R -2\Lambda  + 16\pi G {\cal L}_{\rm matter}\right) \ ,
\end{equation}
where $\Lambda=-6/\ell^2$ is the cosmological constant, $g_m$ denotes the determinant of the metric tensor
and $M,N,K,L \in \{0,1,2,3,4\}$. 
The Lagrangian density for the matter fields ${\cal L}_{\rm matter}$ reads~:
\begin{equation}
{\cal L}_{\rm matter}=  - 
\left(\partial_M\Phi\right)^{\dagger} \partial^M \Phi - U( |\Phi|)  \ \ ,
\end{equation}
where $\Phi$ denotes a doublet of scalar fields~: $\Phi = (\phi_1,\phi_2)^t$. We choose a self-interaction potential
of the form
\be
\label{susy_pot}
        U( |\Phi|) = m^2 \eta^2 \left(1 - \exp\left( - \frac{|\Phi|^2}{\eta^2} \right) \right) = m^2 |\Phi|^2 - m^2\eta^2\sum\limits_{n=2}^{\infty}
        \frac{(-1)^n}{n!} \left(\frac{|\Phi|^{2}}{\eta^2}\right)^n\ee
where $m$ denotes the scalar boson mass and $\eta$ is an energy scale. This type of potential was used previously in the construction
of spherically symmetric boson stars in aAdS \cite{Hartmann:2012wa,Hartmann:2012gw}. This potential is a continuous
version of a scalar field potential that appears in gauge-mediated Supersymmetry (SUSY) breaking \cite{cr,ct}. 
In the second equality in (\ref{susy_pot})
we have separated the mass term from the self-interaction terms in the potential. 

The coupled field equations for matter and gravity  are obtained from the variation of the
action with respect to the matter and metric fields respectively, leading to
\begin{equation}
\label{full_eqs}
\frac{1}{\sqrt{-g_{\rm m}}}\partial_M (\sqrt{-g_m} \partial^M \Phi) = \frac{\partial U}{\partial \Phi^{\dagger}} \ , \  
G_{MN} + \Lambda g_{MN}=8\pi G T_{MN} \ ,
\end{equation}
where $T_{MN}$ is the energy-momentum tensor given by 
\begin{equation}
 T_{MN}=g_{MN} {\cal L}_{\rm matter} - 2\frac{\partial {\cal L}_{\rm matter}}{\partial g^{MN}} \ .
\end{equation}

Our aim is to construct rotating  boson stars in aAdS.
In general, such a solution
would possess two independent angular momenta associated to the two orthogonal planes of rotation present in (4+1)-dimensional
space-time. Here we  restrict to the case of equal angular momenta. The symmetry of the solution can then be enhanced and the
Ansatz for the metric reads \cite{Hartmann:2010pm}
\begin{eqnarray}
\label{metric}
ds^2 & = & -b(r) dt^2 + \frac{1}{f(r)} dr^2 + g(r) d\theta^2 + h(r)\sin^2\theta \left(d\varphi_1 - 
W(r) dt\right)^2 + h(r) \cos^2\theta\left(d\varphi_2 -W(r)dt\right)^2 \nonumber \\
&+& 
\left(g(r)-h(r)\right) \sin^2\theta \cos^2\theta (d\varphi_1 - d\varphi_2)^2 \ ,
\end{eqnarray}
where $\theta$ runs from $0$ to $\pi/2$, while $\varphi_1$ and $\varphi_2$ are 
in the range $[0,2\pi]$.
The corresponding space-times  possess two rotation planes at $\theta=0$ and $\theta=\pi/2$ and the isometry
group is $\mathbb{R}\times U(2)$. 
The metric above still leaves the diffeomorphisms related to the definitions of the radial variable $r$ unfixed. 
We will later fix this  freedom by choosing $g(r)=r^2$. 
For the scalar doublet we choose
\be
          \Phi(r,t) = \phi(r) e^{i \omega t} \hat \Phi
\ee
where $\hat \Phi$ is a doublet of unit length that depends on the angular coordinates.
For solutions without orbital angular momentum we choose 
\begin{equation}
\label{hatphi_non}
 \hat \Phi = (1,0)^T  \ ,
\end{equation}
while for
rotating solutions we have \cite{Hartmann:2010pm}
\be
\label{hatphi_rot}
\hat \Phi = \left(\sin \theta e^{i \varphi_1},\cos \theta e^{i \varphi_2}  \right)^T \ .
\ee  
In this latter case the boson star solution possess only one single Killing vector field. While the metric (\ref{metric})
has three commuting Killing vector fields $\partial_t$, $\partial_{\varphi_1}$, $\partial_{\varphi_2}$ \cite{Hartmann:2010pm}, 
the scalar doublet of the rotating solution with (\ref{hatphi_rot}) is only invariant 
under one possible combination of these vectors \cite{Dias:2011at}, namely under
\begin{equation}
 \partial_t - \omega\left(\partial_{\varphi_1} + \partial_{\varphi_2}\right) \ .
\end{equation}

The mass $M$ and angular momenta $J_1=J_2\equiv J$ can be found by using the appropriate Komar expressions.
They can equally be read off from the asymptotic expansion of the matter and metric fields \cite{Hartmann:2010pm}. 
These are given by
 \be
\label{inf1}
 f(r \gg 1) = 1 +  \frac{r^2}{\ell^2} + \frac{\tilde{{\cal M}}}{r^2} + O(r^{-4}) \ \ , \ \ b(r\gg 1) = 1 + \frac{r^2}{\ell^2} + \frac{{\cal M}}{r^2} + O(r^{-4}) 
 \ee
 \be
\label{inf2}
 h(r\gg 1) = r^2 + O(r^{-2}) \ \ , \ \ W(r\gg 1) = \frac{{\cal J}}{r^4}  \ .
 \ee
At the same time the scalar field function behaves like
\be
\label{inf3}
         \phi(r \gg 1)= \frac{\phi_0}{r^{\Delta}} \ \ , \ \ \Delta = 2 + \sqrt{4 + \ell^2 m^2} \ 
\ee
in the case of asymptotically AdS ($\Lambda < 0$) and 
\begin{equation}
\label{bc2a}
 \phi(r \gg 1)\sim \frac{1}{r^{\frac{3}{2}}} \exp\left(-\sqrt{1-\omega^2}r\right) + ...
\end{equation}
in the case of asymptotically flat solutions ($\Lambda=0$), respectively.

The parameters $\tilde{{\cal M}}$, ${\cal M}$,  and ${\cal J}$ have to be determined numerically and then give 
the mass $M$ and angular momentum $J$ of the solutions~: 
\begin{equation}
 M =  \frac{\pi}{8G} \left(\tilde{{\cal M}} -4{\cal M}\right) \ \ , \ \ 
J  =\frac{\pi}{ 4 G} {\cal J} \ \ .
\end{equation}
Interestingly, the Ansatz here is such that the angular momenta are proportional to the
Noether charge $Q$ associated to the global $U(1)$ symmetry $\Phi \rightarrow e^{i\alpha} \Phi$ 
of the solutions. The locally conserved Noether current reads
\begin{equation}
 J^M = -i\left(\Phi^{\dagger} \partial^M \Phi - \partial^M \Phi^{\dagger} \Phi\right) \ \ , \ \ M=0,1,2,3,4 \ .
\end{equation}
with the globally conserved Noether charge given by
\begin{equation}
 Q= -\int \sqrt{-g_m} J^0 d^4 x  \ . 
\end{equation}
As was shown in \cite{Hartmann:2010pm} the angular momentum $J$ is related to $Q$ by $J=Q/2$.

Inserting the Ans\"atze into the equations of motion (\ref{full_eqs}) leads to a set of coupled non-linear ordinary differential
equations which are very involved, this is why we do not present them here. However, these can be easily reconstructed by applying the variational
principle to the reduced 1-dimensional action which can be found from (\ref{action}) by using the explicit expression
\begin{equation}
\sqrt{-g_m} = r^2 \sin \theta \cos \theta \sqrt{\frac{bh}{f}}   \ , 
\end{equation}
\begin{equation}
 R= \frac{f}{b} \left( b' g' + \frac{g}{2h} b' h' + \frac{b}{2g} (g')^2 + \frac{b}{h} g' h' + \frac{1}{2}gh(W')^2
 + \frac{2b}{f} \left(4 - \frac{h}{g}\right)  \right)
 \end{equation}
 and
\begin{equation}
 { \cal L}_{\rm matter} = -f (\phi')^2 - \phi^2\left(\frac{2}{g} + \frac{1}{h}\right) + \left(W + \omega\right)^2 \frac{\phi^2}{b} - U(\phi)  
\end{equation} 
Here and in the following the prime  denotes the derivative with respect to $r$. 
Fixing the metric gauge by $g(r)=r^2$ leads to a system of four coupled equations for
the functions $f$, $b$, $W$ and $\phi$ plus a constraint. The equation for $f$ is first order, while all the other equations are second order.
the constaint is fullfilled as a consequence of the equations.
Accordingly, nine  boundary conditions have to imposed.

Four of these conditions  follow from the regularity of the solution  at the  origin 
\begin{equation}
f(0) = 1 \ \ , \ \ b'(0)=0 \ \ , \ \  \tilde{h}'(r=0)=0 \ \ , \ \ W'(0)=0 \ \ , \ \  \phi(0)=0 
\end{equation}
where, for convenience, we set  $\tilde{h}(r)\equiv h(r)/r^2$. 
On the conformal boundary we impose the behaviour given in (\ref{inf1}), (\ref{inf2}) and
(\ref{inf3}). 

\section{Boson stars without self-interaction and massive oscillons}
In the case of a fixed AdS background ($G\equiv 0$) and without a scalar field potential it was shown that the scalar
field equation possesses so-called {\it oscillon solutions} \cite{Bizon:2011gg} which act as a basis in
AdS. These solutions have been generalized to higher dimensions \cite{Jalmuzna:2011qw}  as well as to the rotating case \cite{Dias:2012tq}.
Here we will show that this result can be extended
to include the mass of the scalar field. Taking only the mass term in (\ref{susy_pot}) into account, i.e. neglecting 
any non-linear terms in the potential, the scalar field
equation in a fixed AdS background reads:
\begin{equation}
      \partial_x\left(x^3 f \partial_x \phi\right) - x^3 \phi \left(\frac{3\delta_{j1}}{x^2} - \frac{\ell^2\omega^2}{f}\right) -\ell^2 m^2 x^3\phi  = 0 \ \ , \ \   j=0,1 \ \ , \ \ 
      f(x) = 1 + x^2 \ \ , \ \ x:=\frac{r}{\ell} 
\label{scalareq_both}
      \end{equation}
where $j=0$ corresponds to the non-rotating and $j=1$ to the rotating case, respectively. 
This equation has solutions in terms of {\it massive oscillons} which can be given as follows
\be
\phi_k(x)=c_k x^j (1+x^2)^{-2-\alpha-j/2} \phantom{A}_2F_1\left(\frac{4-\ell\omega+j}{2} 
+\alpha,\frac{4+\ell\omega+j}{2} +\alpha;3+2\alpha ,\frac{1}{1+x^2}\right) \ , \ \ k=0,1,2,3....
\label{basis}
\ee
with 
\be
\alpha=-1+\sqrt{1+\ell^2 m^2}    \ .
\ee
The $\phi_k(x)$ act as a basis in the sense that every solution of the equation (\ref{scalareq_both}) 
can be written as a linear superposition
of the basis functions (\ref{basis}). 
Since we want the hypergeometric function $\phantom{A}_2F_1$ to behave regular at the origin, we need to require (see e.g. \cite{magnus})
\be
\label{oscillon_spectrum}
\ell\omega=4+j+ 2\alpha + 2k \  \ , \ \ k=0,1,2,.....  \ .
\ee

The $c_k$ in (\ref{basis}) are suitable constant to be fixed by the boundary conditions. Note that the value
of $\omega$ is fixed by the choice of $m^2$, $\ell^2$ and $k$ such that the spectrum of the oscillons does not depend on
$c_k$. This is reminiscent of a zero-mode, i.e. a continuous family of solutions that all have the same energy.
 We show the first three massive 
oscillon solutions with $m^2=2$ for the non-rotating case in Fig.\ref{mo}. We note that the difference between the profiles
of the massless and the massive oscillons is small.
The inlet shows the profiles of a massive
and a massless oscillon for $k=2$, respectively.

\begin{figure}[h]
\begin{center}
\includegraphics[width=10cm]{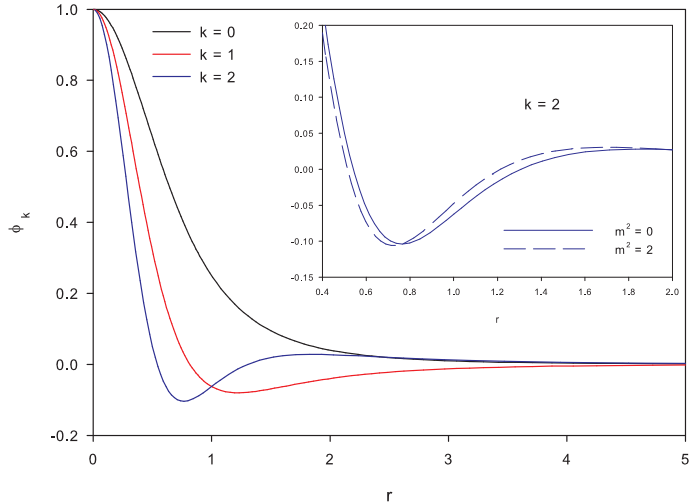}
\caption{\label{mo} We show the first three massive oscillon solutions with normalisation such that $\phi_k(0)=1$ and
for $m^2=2$. The small inlet shows the difference between the massless oscillon ($m^2=0$) and the massive oscillon ($m^2=2$)
for $k=2$. }
\end{center}
\end{figure}

When either the self-interaction of the scalar field and/or backreaction of the space-time is taken into account the equations of motion
become non-linear and the spectrum (\ref{oscillon_spectrum}) will change. We follow the literature and call the resulting solutions 
{\it boson stars} in the following as compared to the analytically given solutions of the linear scalar field equation which we refer
to as {\it massive oscillons}.

\section{Gravitating and self-interacting boson stars}
Considering now the full scalar field potential (\ref{susy_pot}) as well as the case $G\neq 0$ the set of coupled
scalar and Einstein equations can no longer be solved analytically. 
We have hence solved this system of coupled differential equations subject to the appropriate boundary conditions
using an adaptive grid Newton-Raphson scheme \cite{colsys}.
The solutions are constructed with  relative errors  which  are typically on the order of $10^{-7}-10^{-9}$.  

The coupled scalar field and Einstein equations depend on four constants: the gravitational coupling $G$, 
the cosmological constant $\Lambda$ (or the Anti-de Sitter radius $\ell$),
the mass of the scalar field $m$ and the energy scale $\eta$. While in the discussion on 
massive oscillons we kept all constants we find it useful for our numerical calculations to 
redefine the radial variable $r$, the scalar field $\phi$ and the frequency $\omega$
according to
\begin{equation}
 r \rightarrow \frac{r}{m}   \ \ , \ \ \phi \rightarrow \eta\phi \ \ , \ \ \omega \rightarrow m\omega  \ .
 \end{equation}
The equations then depend only on two dimensionless coupling constants given by
\be
            \kappa \equiv 16 \pi G \eta^2 \ \ , \ \ Y \equiv \frac{1}{m^2 \ell^2} \ .
\ee
In principle, we have thus four parameters to vary: $Y$, $\kappa$ and $\omega$ (or equivalently $\phi'(0)$), which take on continuous
values, as well as the number of nodes of the scalar field function, which takes on discrete values. This is a large parameter space and
we have in the following restricted ourselves to specific values of these parameters that we believe capture the main qualitative
features of the model. In particular, we are interested in understanding the deformation of massive oscillons mentioned above.

Using the Ansatz (\ref{hatphi_non}) for the scalar field we recover the known non-rotating  boson stars.
These have been studied in several papers \cite{Astefanesei:2003qy,Hartmann:2012gw,Stotyn:2013yka}.
In this paper, we concentrate on the rotating solutions by using the Ansatz (\ref{hatphi_rot}).
The asymptotically flat case ($Y=0$) has been studied in \cite{Hartmann:2010pm} for both $\kappa=0$ and $\kappa > 0$. 
The specific Ansatz for the scalar field function that leads to solutions with only one Killing vector
was introduced for the first time in that latter paper, where a 6th order
scalar potential was used. The potential (\ref{susy_pot}) used in this paper possesses a mass term as well as a full series 
of powerlike interactions encoded in the exponential. In particular this implies that the scalar field equation becomes non-linear
(even for $G=0$). The non-linearity then leads to the fact that the value of $\phi(0)$ for the non-rotating case
or $\phi'(0)$ in the rotating case, respectively, is no longer a free parameter but fixes the value of $\omega$ (for given values of
$m^2$ and $\ell$).
Hence, the zero-mode in the oscillon spectrum disappears in the non-linear case. 
We have first been interested in the modification of the oscillon spectrum. For this purpose we
have studied a specific solution with fixed $\phi(0)$ or $\phi'(0)$, respectively.

\subsection{Change of the spectrum in the non-linear case}

First, we have studied the modification of the spectrum (\ref{oscillon_spectrum}) for $\kappa\neq 0$, i.e. taking 
backreaction of the space-time into account, but neglecting self-interaction of the scalar field. 
Our results are shown in Fig.\ref{gmo} where we compare the analytically given oscillon spectrum (\ref{oscillon_spectrum}) 
for $\kappa=0$ with the numerically calculated boson star spectrum for $\kappa=0.5$ in the non-rotating case ($j=0$).

\begin{figure}[ht]
\begin{center}
\includegraphics[width=10cm]{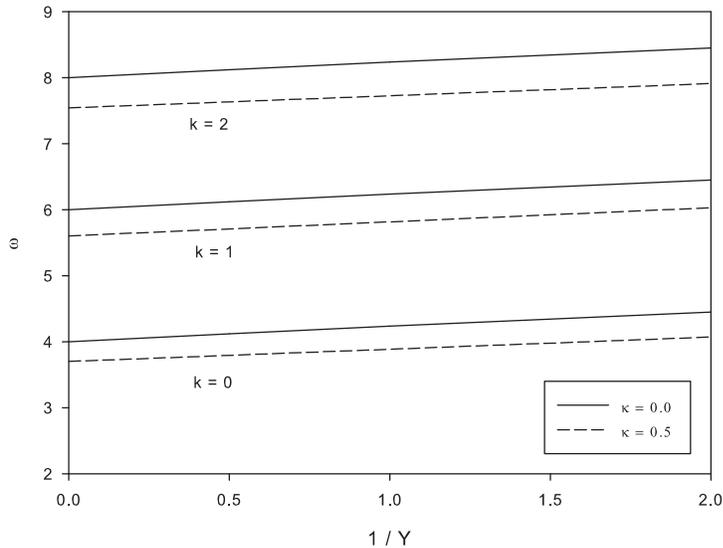}
\caption{\label{gmo} We show the spectrum of the first three boson stars ($k=0,1,2$) for the non-rotating case ($j=0$) as function of $1/Y$ for
$\kappa=0.5$. We compare this to the spectrum of the first three oscillons ($\kappa=0$). Note that the corresponding
scalar field profiles have all $\phi(0)=1$.}
\end{center}
\end{figure}
We clearly observe that $\omega$ decreases for all $k$ when increasing the backreaction of the space-time, however, the qualitative
dependence of the spectrum on the mass of the scalar field is very similar. 

There is one important point to mention here. While the value of $\omega$ is fixed by the choice of the remaining 
(dimensionless) parameters in the model in the linear case this is different for the non-linear case (here the gravitating case, but similar
arguments apply when including the self-interaction of the scalar field).
In the linear case, we can fix the value of $\phi(0)$ at will, the spectrum will not change. However, in the non-linear case the modes
mix and $\omega$ is now depending on the choice of $\phi(0)$. For Fig.\ref{gmo} we have chosen
$\phi(0)=1$. If we had chosen a different $\phi(0)$ the spectrum would change. To state it differently: in the non-linear
case we have a full band of values $\omega\in [\omega_{\rm min}:\omega_{\rm max}]$ for a given value of $1/Y$, while
for the linear case of a massive scalar field in a fixed AdS background one unique value of $\omega$ exists for each $1/Y$.

Our discussion above is done for the non-rotating case. In the rotating case, which we are mainly interested in here, 
we have to fix $\phi(0)=0$ to ensure regularity on the axis of rotation. Hence, we use $\phi'(0)$, the derivative of the
scalar field function at $r=0$, and construct solutions varying this value. Our results are discussed in the following.

\subsection{Mass and charge of rotating boson stars}
In the following we will discuss solutions taking the full potential (\ref{susy_pot}) into account and comparing the cases of a fixed AdS
background ($\kappa=0$) and a back-reacting space-time ($\kappa > 0$), respectively.
We have first studied the dependence of the physical quantities on the 
parameter $\phi'(0)$ that we have mentioned above.
Our results are shown in Fig. \ref{fig_1_more} where we
give the dependence of the charge $Q$ and of the frequency $\omega$ on the parameter $\phi'(0)$ for different values of $Y$.

\begin{figure}[ht]
\begin{center}
\subfigure[][ $\omega$ in dependence on $\phi'(0)$]{\label{omegaphi}
\includegraphics[width=8cm]{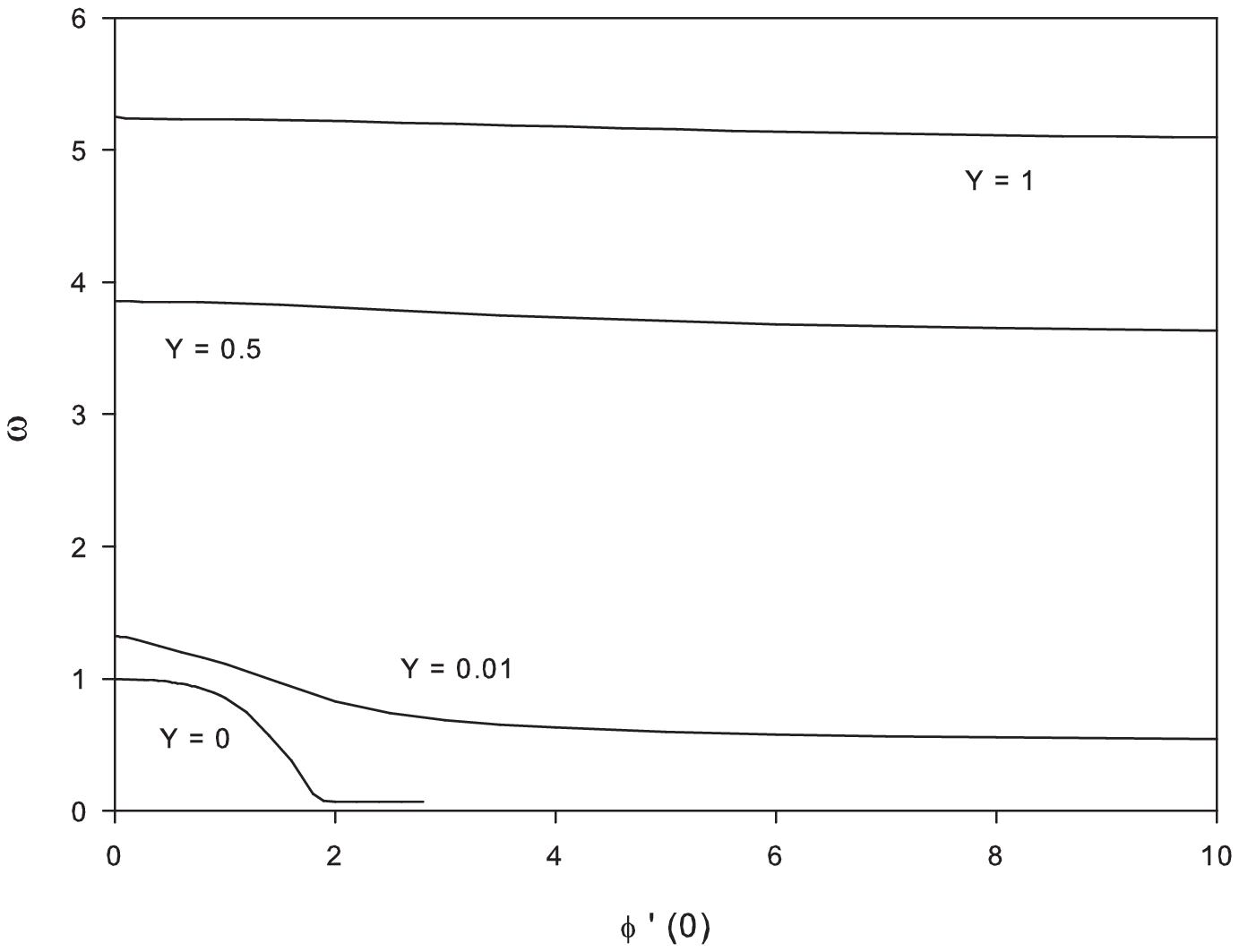}} 
\subfigure[][$Q$ in dependence on $\phi'(0)$]{\label{qphi}
\includegraphics[width=8cm]{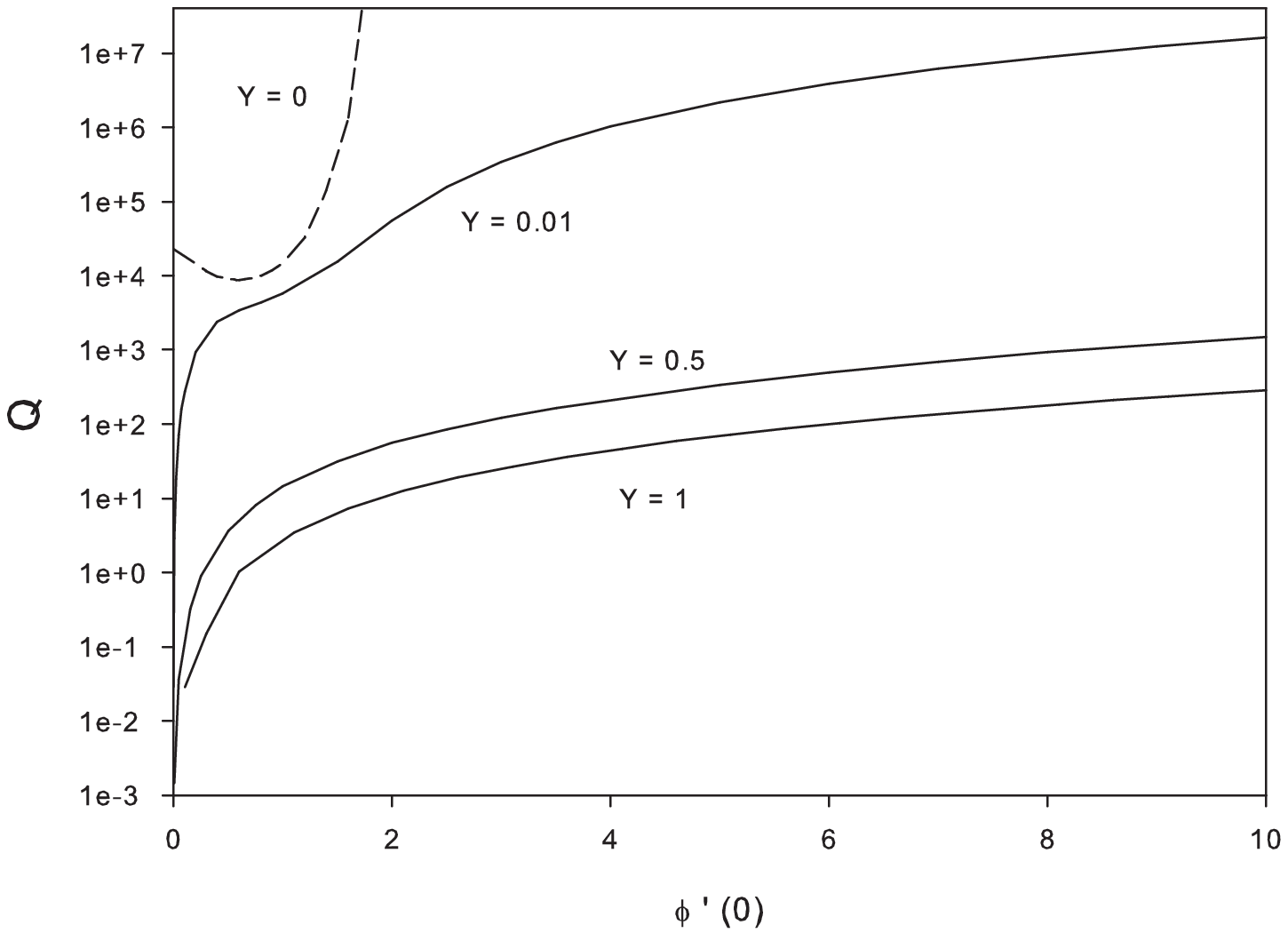}}
\end{center}
\caption{\label{fig_1_more} We show the dependence of the frequency $\omega$ (left) and of the charge $Q$ (right) on $\phi'(0)$ for $\kappa=0$. }
\end{figure}

\begin{figure}[ht]
\begin{center}
\subfigure[][ $\kappa=0$]{\label{probe_limit}
\includegraphics[width=8cm]{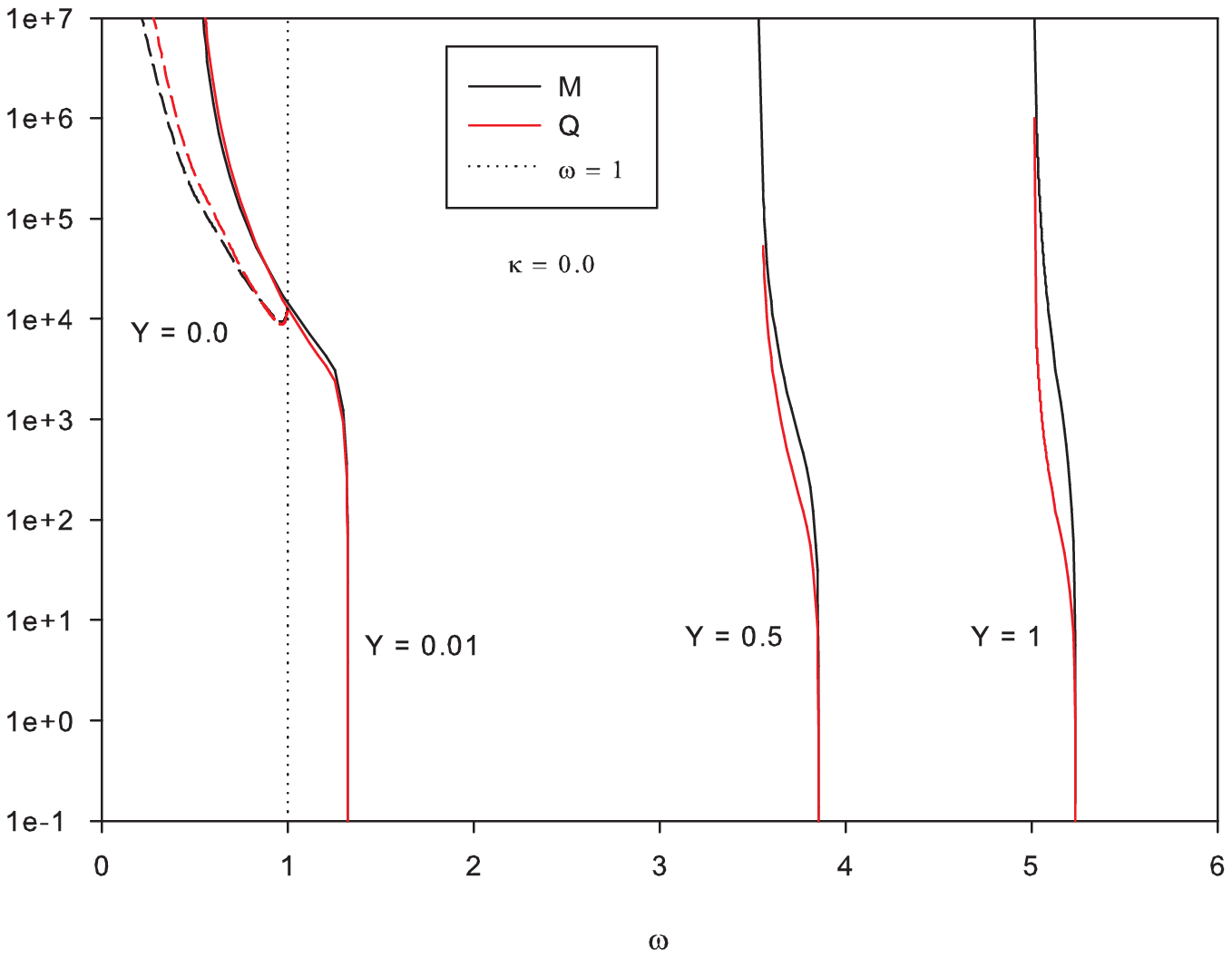}} 
\subfigure[][$\kappa=1$]{\label{kappa1}
\includegraphics[width=8cm]{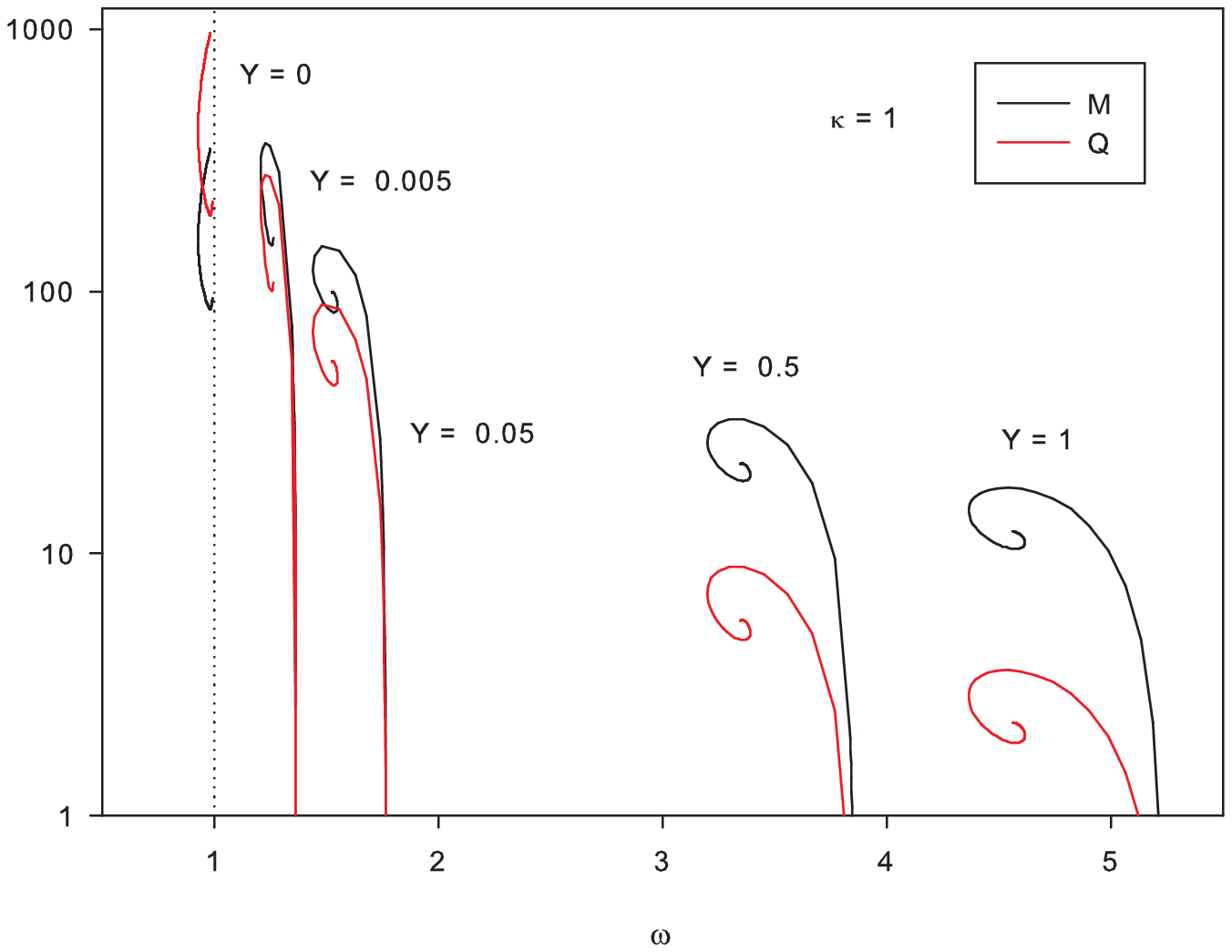}}
\end{center}
\caption{\label{fig_4} We show the dependence of  the mass $M$ (black) and charge $Q$ (red) of rotating boson stars in a fixed AdS background ($\kappa=0$, left) and in a back-reacting space-time ($\kappa=1$, right) 
for different values of $Y$ in dependence on the frequency  $\omega$. The corresponding values for
the asymptotically flat case ($Y=0$) are supplemented (dashed black and red lines, respectively). Note that the angular momentum is given by $J=Q/2$.}
\end{figure}

We observe (in agreement with the discussion above) that for each choice of the parameters $\kappa$ and  $Y$ a family of solutions in the frequency $\omega$ 
can be constructed. The frequency, however, can only take values inside a given interval $\omega\in[\omega_{\rm min}:\omega_{\rm max}]$.
For $\omega$ close to the maximal frequency $\omega_{\rm max}$ the scalar field is spread over a large interval of the coordinate $r$ and 
$\phi'(0)$, the derivative of the scalar field function at the origin is very small. 
In the other limit $\omega\rightarrow \omega_{\rm min}$ the scalar field becomes very strongly concentrated 
at the origin with $\phi'(0)$ becoming very large. In this limit, the boson stars have hence large central densities. 

Furthermore, the case $Y=0$ is qualitatively different from the case $Y\neq 0$. 
Let us first discuss the case $Y > 0$. Here solutions with large values of $\phi'(0)$ can be constructed easily.
The mass and charge increase regularly when increasing $\phi'(0)$. In the limit $\phi'(0) \to \infty$, the term due
to the potential in the equation becomes very small and the value of $\omega$ approaches 
the one of the oscillon.
The limit $\phi'(0) \to 0$ is more subtle. Indeed the AdS vacuum is approached, in particular the mass $M$ and the charge $Q$
tend to zero. 
In the case $Y=0$, $\kappa=0$ the solutions have an upper bound for $\phi'(0) =  \phi'(0)_c$ (we find $\phi'(0)_c \approx 2.8$). 
This has already been observed previously \cite{Hartmann:2010pm} and can be understood when looking at Fig.\ref{qphi}. 
At the approach of $\phi'(0)_c$ the charge $Q$ and hence the angular momenta $J$ diverge. Our numerical analysis also
indicates that in this limit the radius of the boson star becomes large. Since there is no force that
counteracts the centrifugal force in this case the solution ceases to exist. Note that this is different
in AdS space-time, where an attractive force counterbalances the rotation and hence solutions with much larger
$\phi'(0)$ exist.

In order to further interpret the results it is also useful to study the mass and the charge as functions of the frequency $\omega$.
This is shown in Fig.\ref{fig_4} for a fixed AdS background $(\kappa=0$) as well as for a back-reacting 
space-time ($\kappa=1$). 
In a fix AdS background both $M$ and $Q$ 
tend to zero in the limit $\omega\rightarrow \omega_{\rm max}$, while  in the limit $\omega\rightarrow \omega_{\rm min}$  both mass and charge tend to infinity independent of the choice of $Y$. 
As far as the numerical values of $\omega_{\rm min}$ and $\omega_{\rm max}$ are concerned we find that
$|\omega_{max}-\omega_{min}|$  decreases with increasing $Y$. 
Also note that for small values of $Y$ the solutions
exist for $\omega < 1$. At $\omega=1$ our results suggest that the mass and charge of the solutions
in an AdS background have a similar mass and charge than the solutions in flat space-time.
Furthermore, our results clearly suggest
that the value of $\omega_{\rm max}$ is independent
of $\kappa$ and that at this value both the mass and the charge of the solutions tend to zero.
Decreasing $\omega$ from this maximal possible value we find that a spiraling behaviour typical for
boson star solutions appears. The first branch of solutions exists down to $\omega_{\rm min}$ and then extends
backwards in $\omega$ to form a spiral. On the first branch the solutions reach their maximal possible mass
and charge. These values increase with decreasing $Y$. Furthermore we find that the extend in $\omega$ in
which the solutions exist decreases with decreasing $Y$. Our numerical results also suggest that these qualitative
features are quite generic and do not depend much on the actual value of $\kappa  > 0$.

\begin{figure}[ht]
\begin{center}
\includegraphics[width=7cm,angle=270]{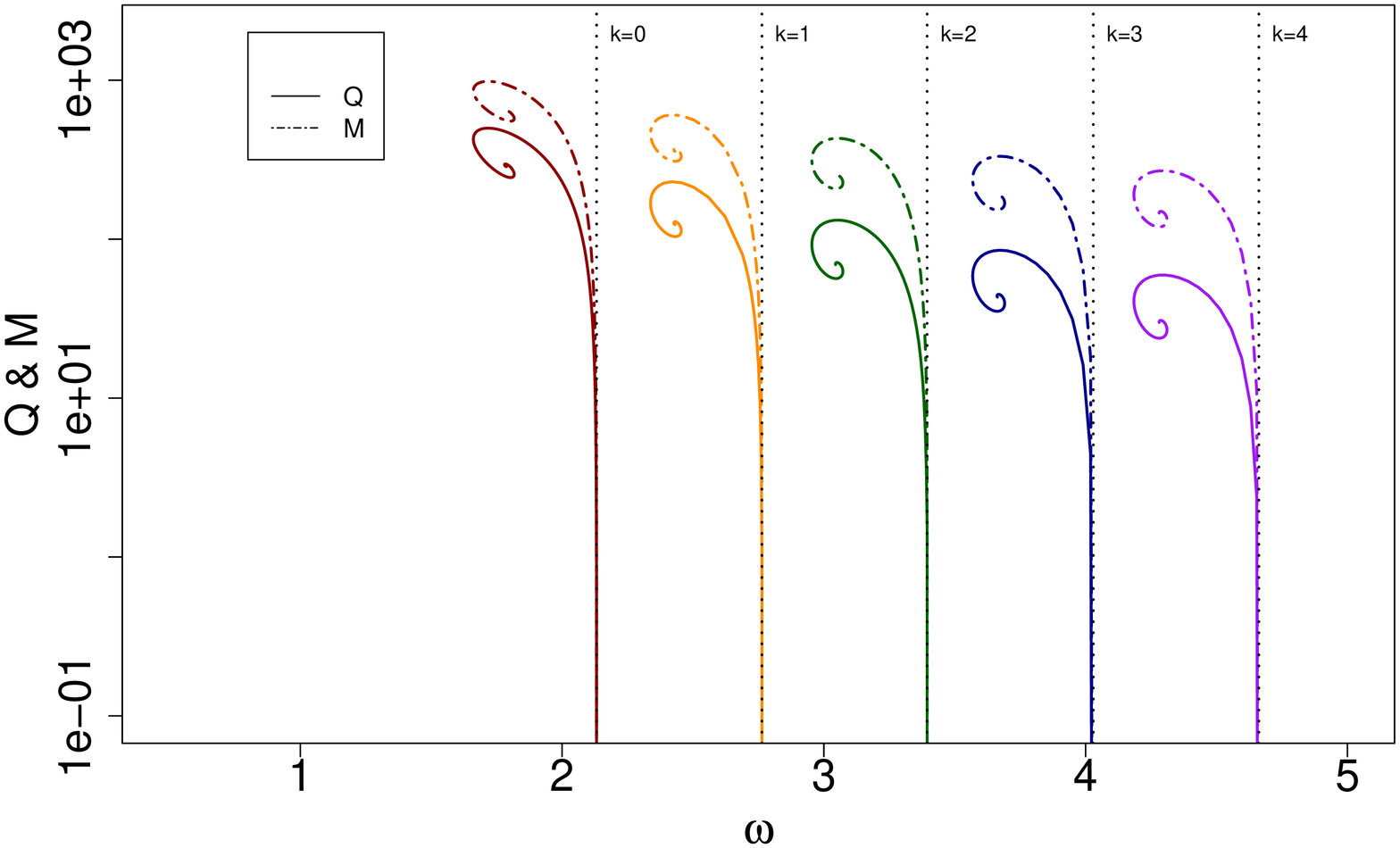}
\end{center}
\caption{\label{excited} We show the mass $M$ (dashed) and the charge $Q$ (solid) as function of $\omega$ for
fundamental ($k=0$) and radially excited rotating boson stars with $k=1,2,3,4$ nodes of the scalar field function.
Here $\kappa=0.1$ and $Y=0.1$.}
\end{figure}

Our above discussion on oscillons suggests that radially excited rotating boson stars should exist.
These solutions possess a number $k$ of nodes of the scalar field function.
We have constructed these solutions numerically for $k=1,2,3,4$ and present the results
in Fig.\ref{excited} for $\kappa=Y=0.1$. We give the mass and charge of the radially excited solutions and 
compare them to 
the fundamental solution with no nodes of the scalar field function ($k=0$). All solutions have mass and charge
tending to zero at the maximal value of the frequency $\omega$. In agreement with our results on oscillons
we find that the maximal frequency $\omega_{\rm max}$ increases with the number of nodes. 
The maximal value of the mass and charge is also a decreasing function of the number of nodes. This implies
as well that the maximal possible
angular momentum $J=Q/2$ decreases with increasing node number. Hence, the radially excited boson stars cannot have angular momentum
as large as the fundamental ones.

\begin{figure}[h!]
\begin{center}
\subfigure[ ][$\kappa=0$]{\label{MJkappa0}
\includegraphics[width=8cm]{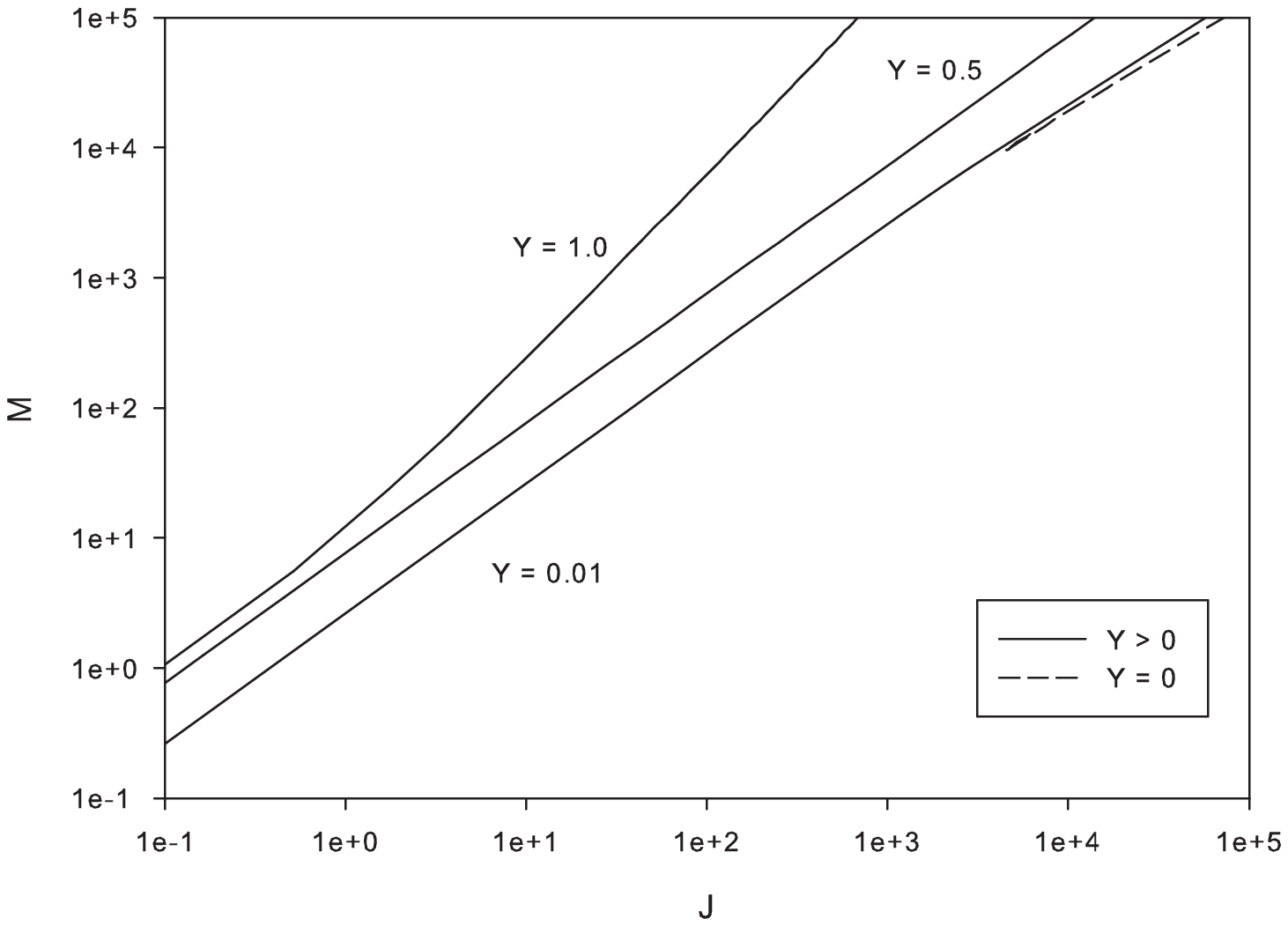}} 
\subfigure[][$\kappa=0.05$]{\label{MJkappa005}
\includegraphics[width=8cm]{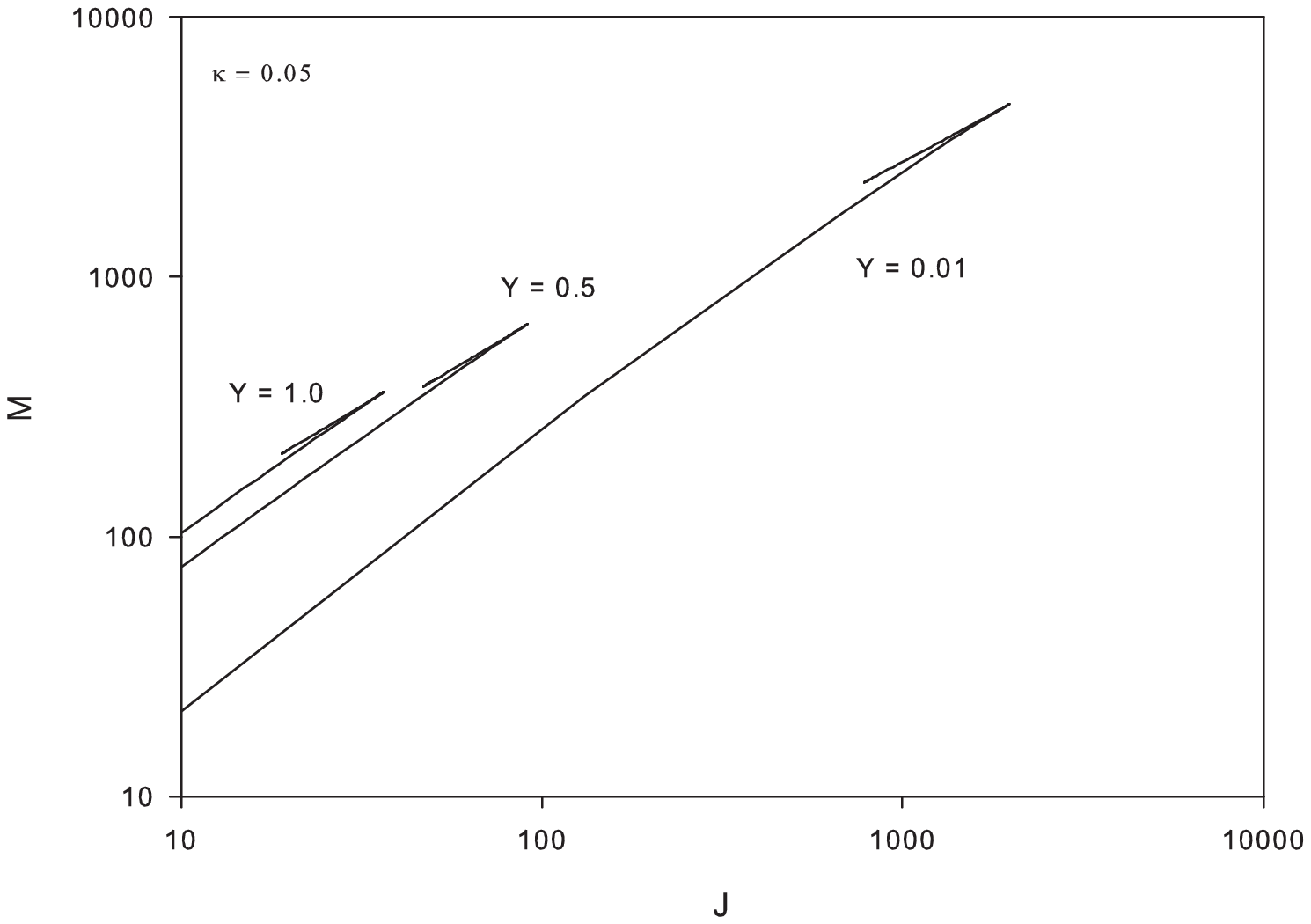}} 
\end{center}
\caption{\label{MJ} 
We show the values of the mass  $M$ in dependence on the angular momentum $J$ for different values of $Y$ and in
a fixed AdS background ($\kappa=0$, left) and in a back-reacting space-time ($\kappa=0.05$, right).
Note that all curves extend to $M=J=0$ which corresponds to the AdS vacuum.}
\end{figure}

Now using the knowledge that AdS is linearly stable \cite{Ishibashi:2004wx} we can say something about the stability of the solutions.
Since catastrophe theory tells us that the stability of solutions remains the same on given branches and changes
only at cusps, we can conclude that the branch of rotating boson stars connected to the AdS vacuum ($M=J=0$) is
stable. For $\kappa=0$ no further branch exist, hence rotating boson stars in a fixed AdS background are linearly stable.
For $\kappa > 0$  a second branch exists that joins the first one at a cusp. This branch is not connected to the AdS vacuum and
we would hence expect the solutions on the second branch (with higher mass $M$ for given $J$ as compared to the first branch) to be
unstable.

\begin{figure}[h!]
\begin{center}
\includegraphics[width=10cm]{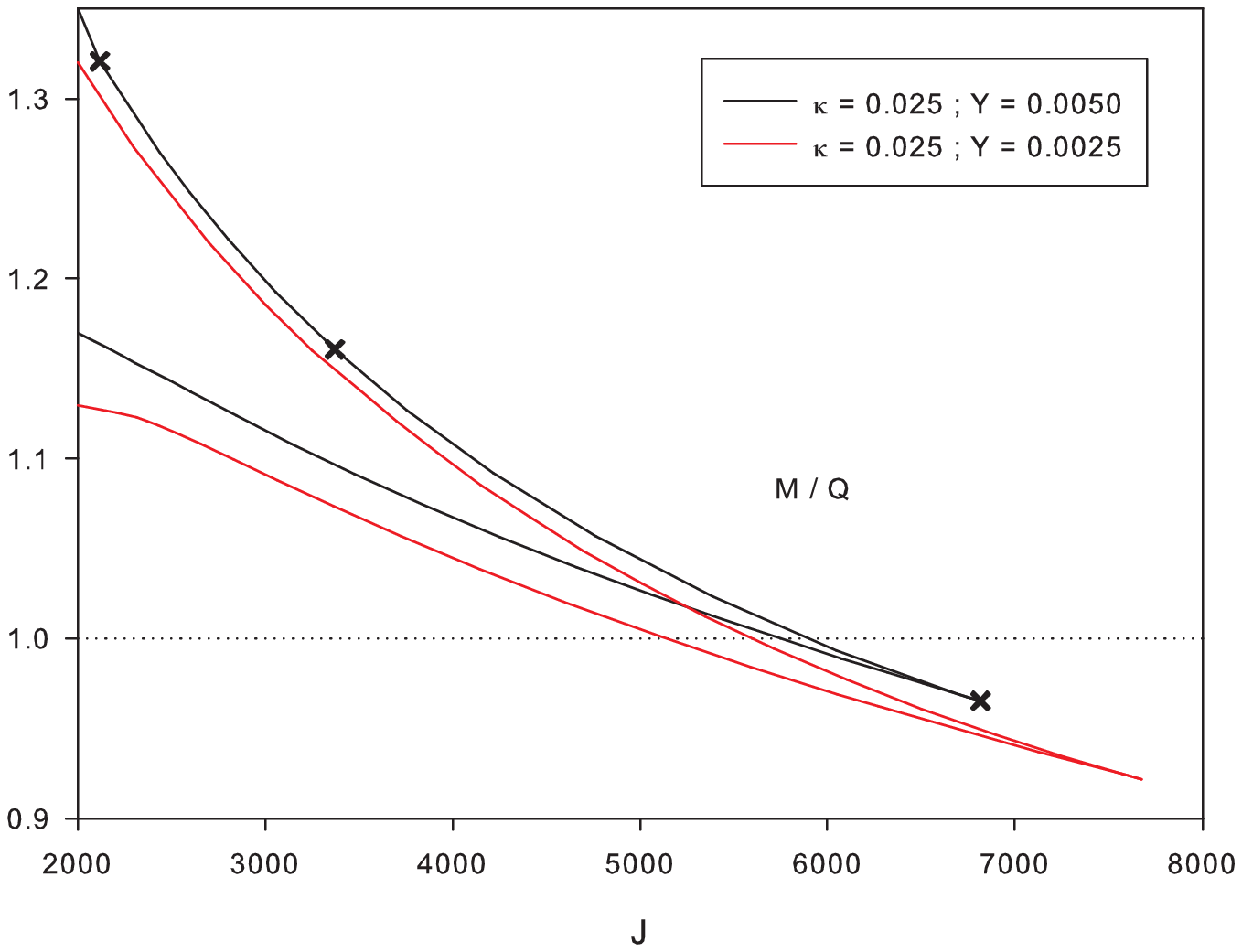}
\end{center}
\caption{\label{stability2} 
We show the values of the mass over the charge $M/Q$ 
in dependence on the angular momentum $J$ for $\kappa=0.025$ and two different values of $Y$. The 
three crosses in the case $Y=0.0050$ (black) correspond to the solutions shown in Fig.\ref{ergo_functions}.
The solution marked by the cross at the smallest value of $J$ possesses an ergoregion (see Fig.\ref{ergo_functions}
for more details).}
\end{figure}

\begin{figure}[h!]
\begin{center}
\subfigure[$-g_{tt}$]{\label{g_tt}
\includegraphics[width=8cm]{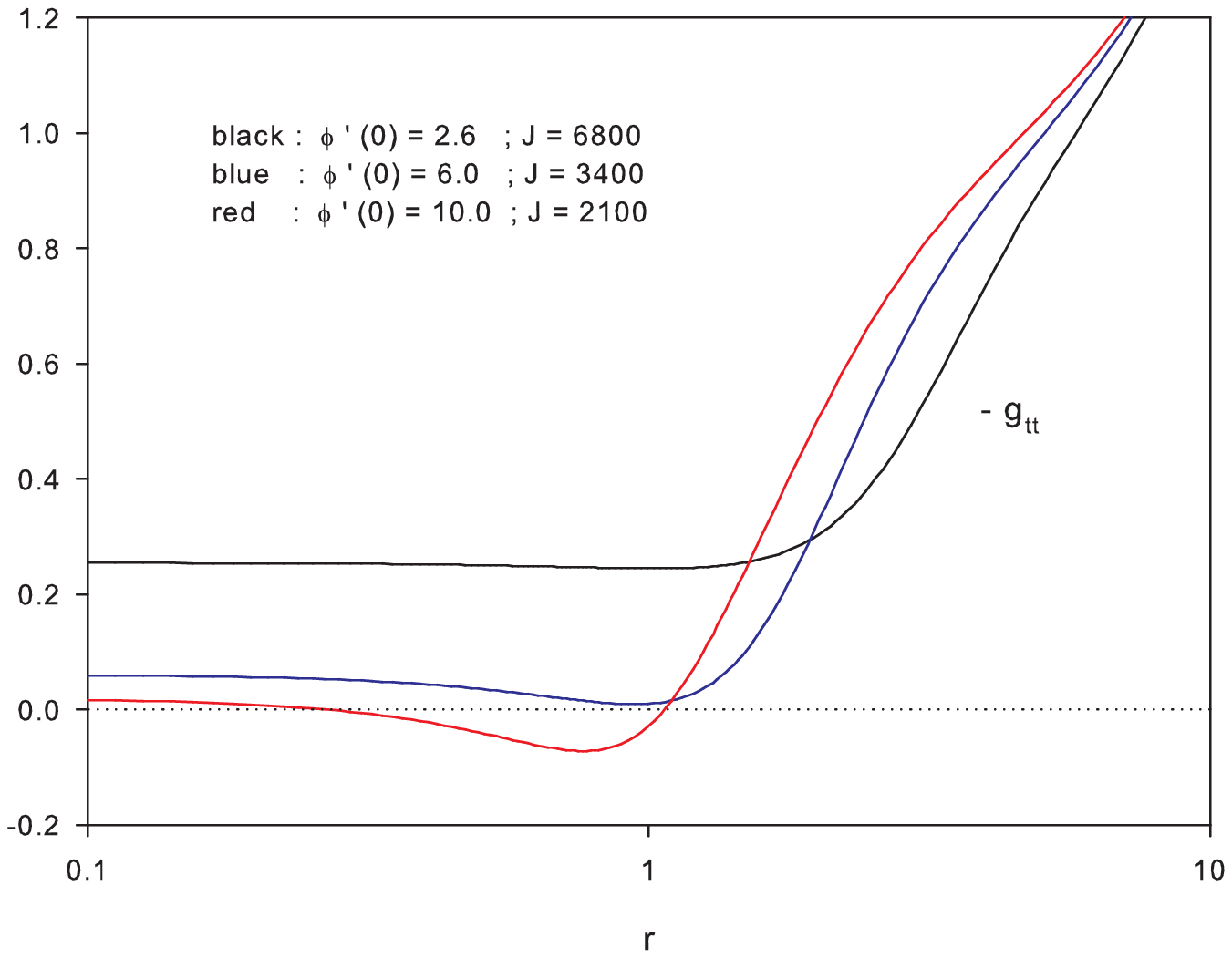}}
\subfigure[$f(r)$ and $b(r)$]{\label{ergofb}
\includegraphics[width=8cm]{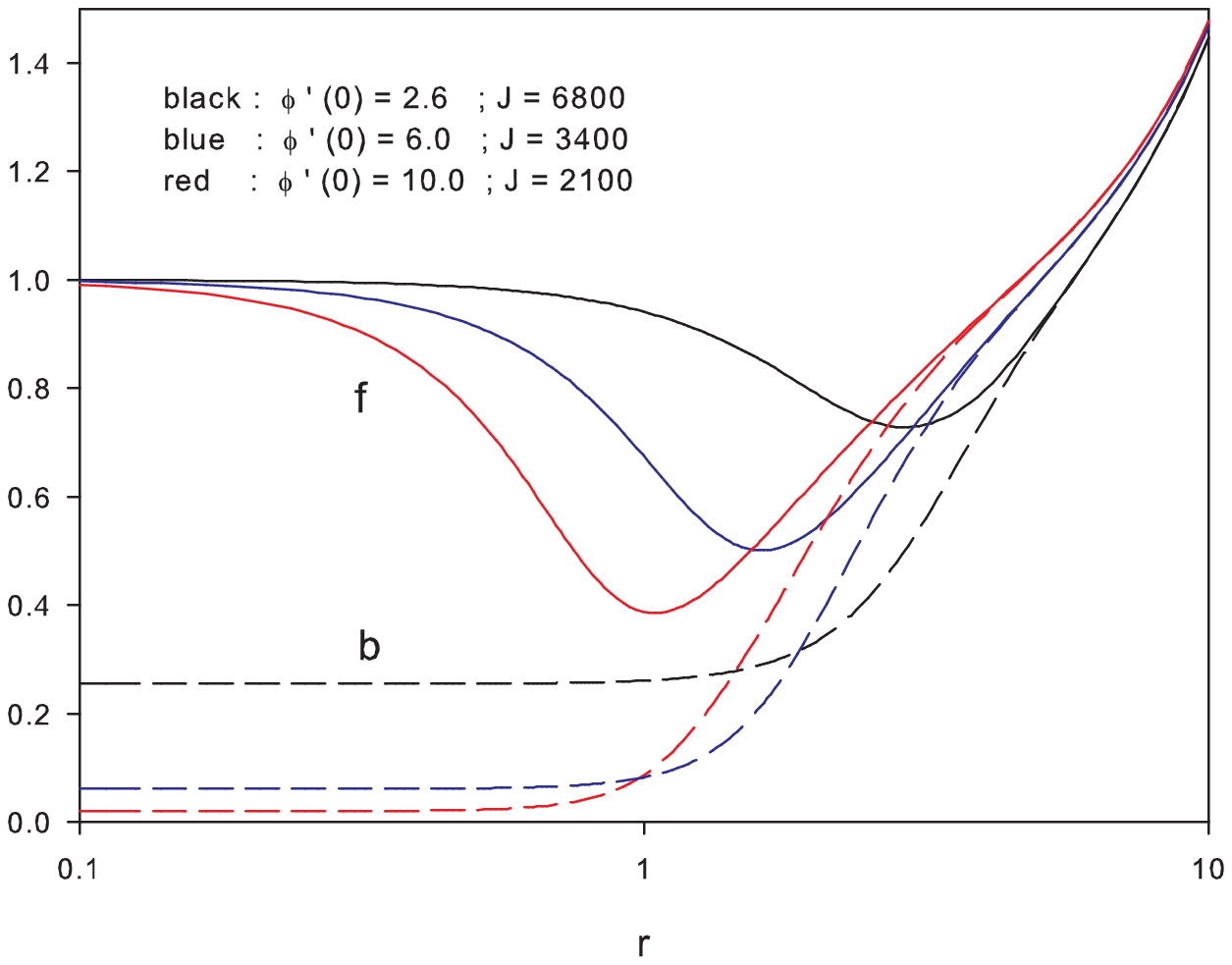}} 
\subfigure[$h(r)/r^2$]{\label{ergoh}
\includegraphics[width=8cm]{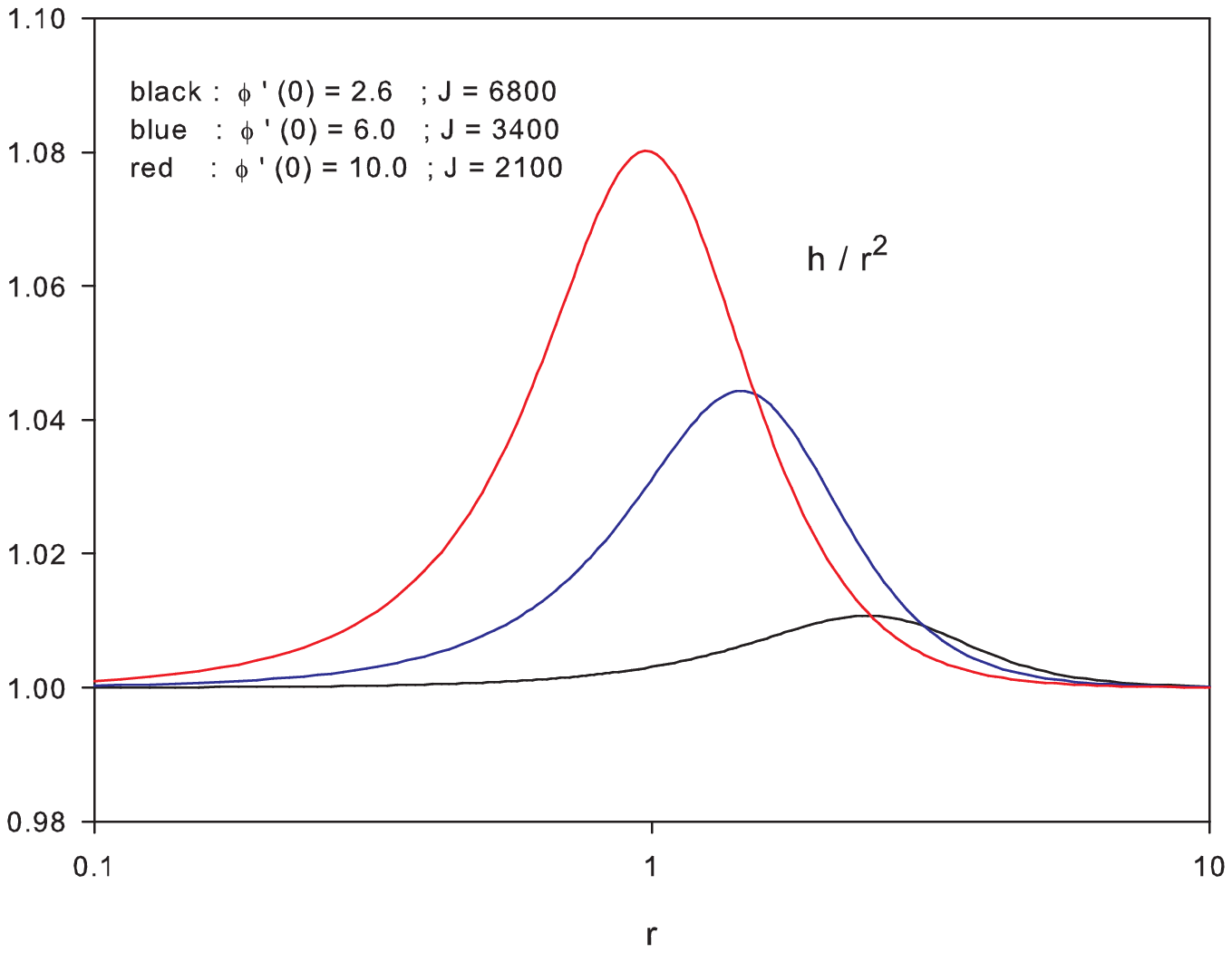}\label{ergoh}} 
\subfigure[$W(r)$ and $\phi(r)$]{\label{ergoWphi}
\includegraphics[width=8cm]{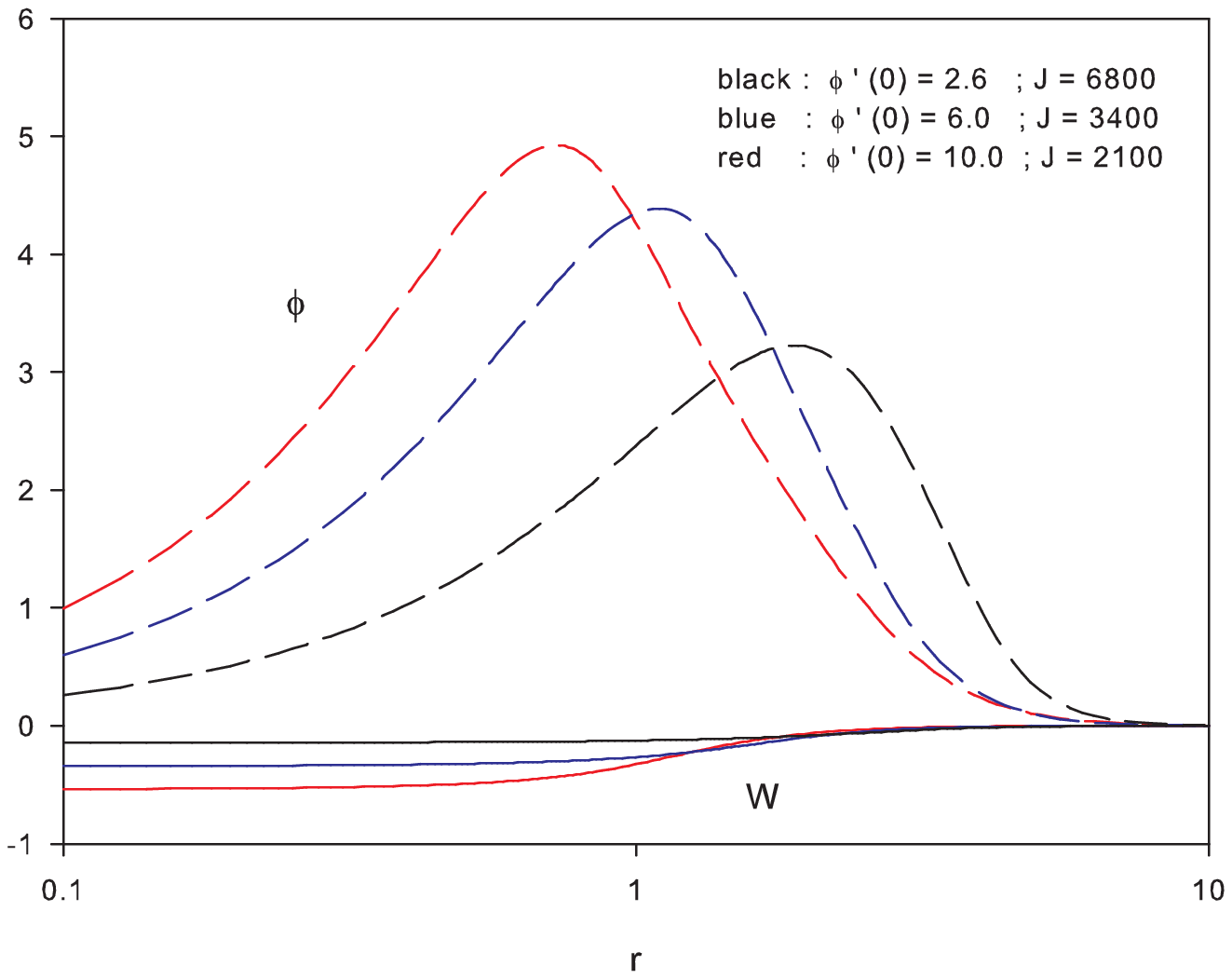}}
\end{center}
\caption{\label{ergo_functions} We show the profiles of $g_{tt}$ as well as the metric functions $f(r)$, $b(r)$, $h(r)$, $W(r)$ and the scalar field function $\phi(r)$ for the solutions marked by the crosses in Fig.\ref{stability2}
}
\end{figure}

\begin{figure}[h!]
\begin{center}

\subfigure[][fundamental solutions ]{\label{ergo_detail1}
\includegraphics[width=8cm,angle=270]{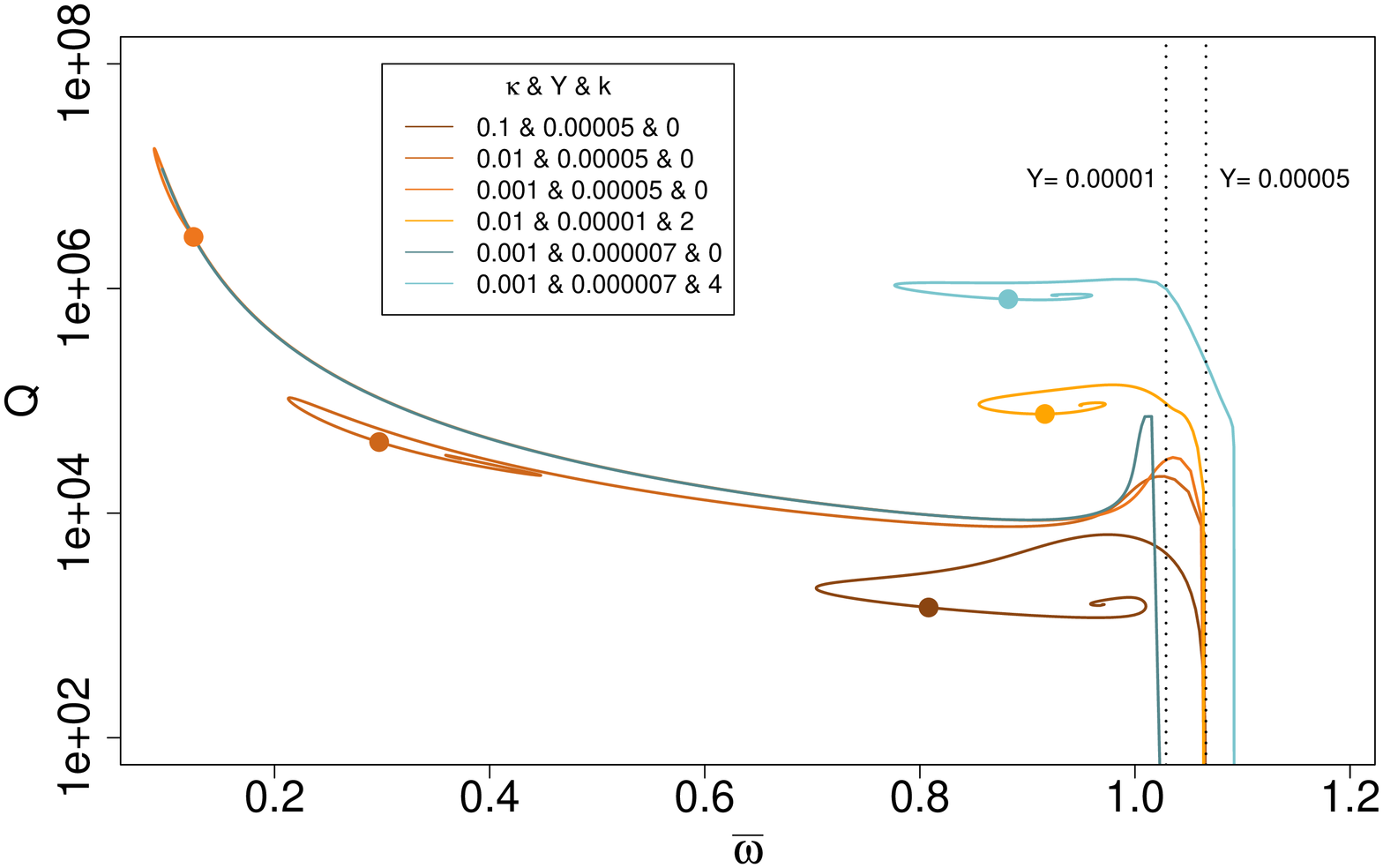}} 
\subfigure[][fundamental and radially excited solutions]{\label{ergo_detail2}
\includegraphics[width=8cm,angle=270]{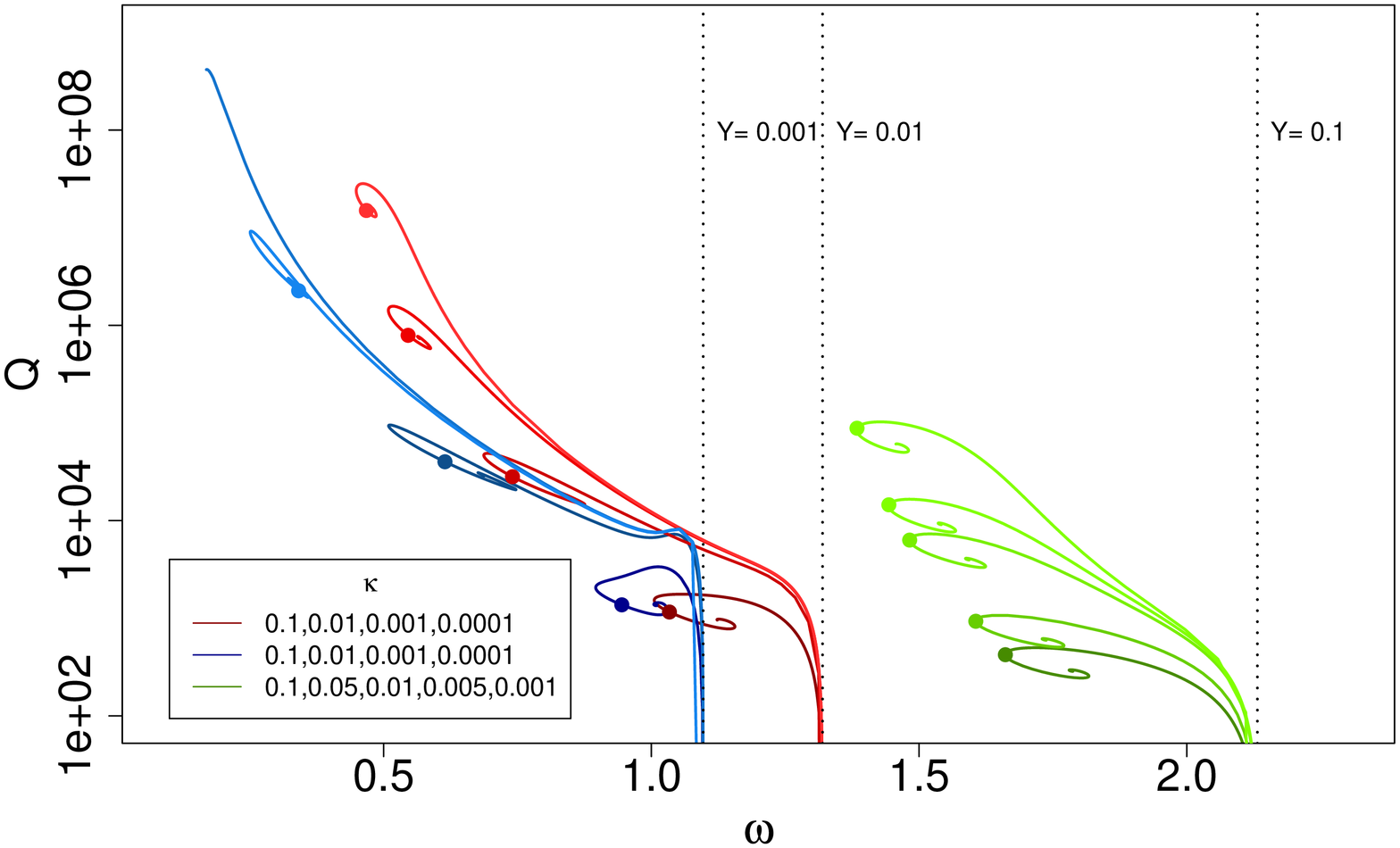}} 
\end{center}
\caption{\label{stability_details} We show the value of the charge $Q$ in dependence on $\omega$
for fundamental rotating boson stars (a) and in dependence on $\bar{\omega}:=(\exp(\omega-1))^3$ for fundamental and radially excited boson stars (b) for different values of $\kappa$ and $Y$. In (a) the value of $Y$ is indicated by the 
vertical line at which the respective branches end with $Q=0$, while the curve
with largest maximal $Q$ corresponds to the smallest value of $\kappa$ (and vice versa). 
Starting from $\omega_{max}$ with $Q=0$ and progressing along the
branch the solutions possess ergoregions from the dot onwards.}
\end{figure}

\section{Stability and ergoregions}

\subsection{Stability arguments from catatrophe theory}

Catastrophe theory \cite{Kusmartsev:1990cr,Tamaki:2010zz} has been widely used to study the stability of boson stars
(for a recent use of the arguments see \cite{Kleihaus:2011sx}). The original idea of catastrophe theory is the study of 
critical points of a given potential which itself depends on a number of independent variables, so-called {\it control variables}.
In addition there are so-called {\it behaviour variables} such that the vanishing of the derivative of the potential with respect to these latter 
variables characterizes the equilibrium situation. In this equilibrium situation the change of stability occurs if the
derivative of one of the control variables with respect to the behaviour variable changes signature. In \cite{Tamaki:2010zz} 
the ``potential'' was chosen to be the total Hamiltonian (which was shown to be proportional to the
gravitational mass of the solution), since the variation of this quantity with respect to the scalar field and the
metric functions, respectively, gives the equations of motion. The control variables were chosen to be the
gravitational coupling as well as the charge of the solution. The behaviour variable was chosen to be $\omega$. 
A very similar startegy can be applied in our case here. The Hamiltonian energy would be equivalently defined and 
can be shown to agree with the gravitational mass $M$. Note that this applies also for the AdS case. This is related to the fact that
in computing the Hamiltonian energy one would have to introduce a boundary counterterm in order to render the integral finite.
However, since it is the energy of the solutions to the equation of motion that is used and the variation of the boundary term does not
contribute to the equations of motion, the gravitational mass read off from the behaviour of the metric functions on the AdS boundary
remains proportional to the Hamiltonian in the equilibrium case.  Below we will use these arguments choosing as a behaviour
variable the angular momentum $J=Q/2$ and as control variables the parameters $\kappa$ and $\phi'(0)$ that describe the solution uniquely.

In Fig.\ref{MJ} we show the dependence of the mass $M$ on the angular momentum $J$ for different values of $Y$ and in a
fixed AdS background ($\kappa=0$) as well as in a back-reacting space-time ($\kappa=0.05$). In both cases the curves
extend all the way down to $M=J=0$ which corresponds to the AdS vaccuum. In the case $\kappa=0.05$ a second branch of solutions
exists that has higher $M$ for a given $J$ and joins the branch connected to the AdS vacuum in a cusp at the maximal possible
value of $J$.

\subsection{Ergoregions}

\begin{figure}[h!]
\begin{center}

\subfigure[][fundamental solutions ]{\label{ergo_detail1}
\includegraphics[width=8cm,angle=270]{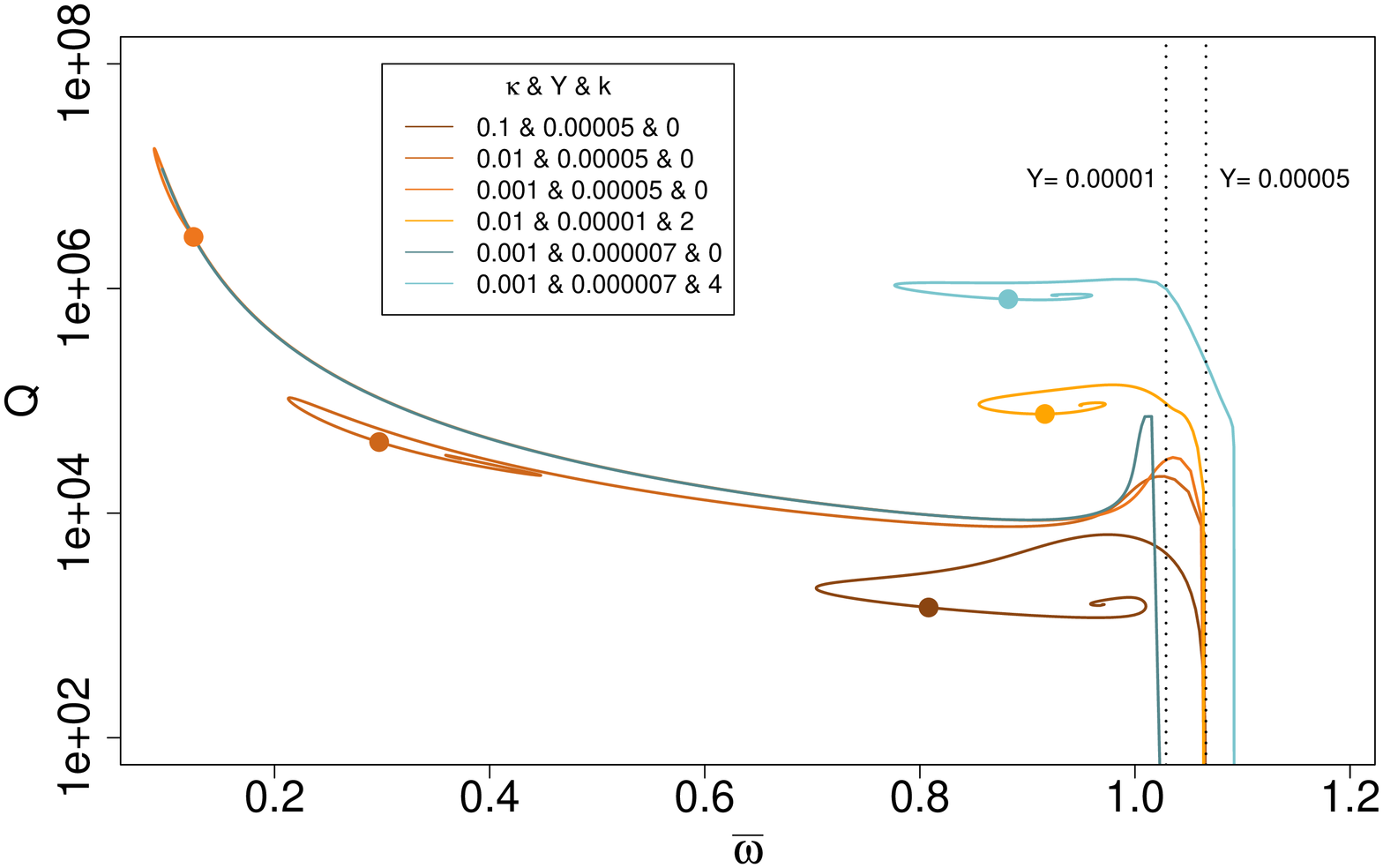}} 
\subfigure[][fundamental and radially excited solutions]{\label{ergo_detail2}
\includegraphics[width=8cm,angle=270]{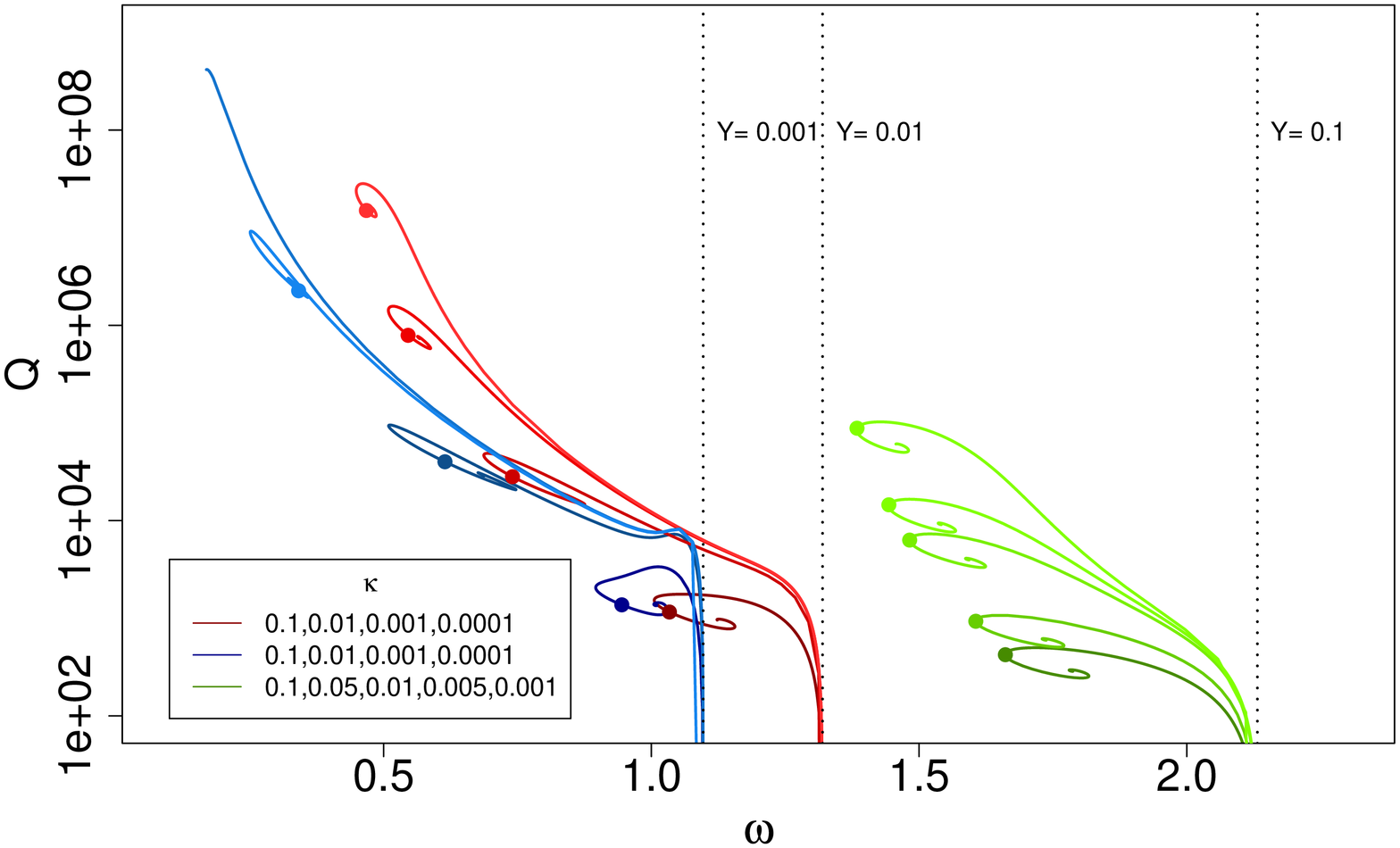}} 
\end{center}
\caption{\label{stability_details} We show the value of the charge $Q$ in dependence on $\omega$
for fundamental rotating boson stars (a) and in dependence on $\bar{\omega}:=(\exp(\omega-1))^3$ for fundamental and radially excited boson stars (b) for different values of $\kappa$ and $Y$. In (a) the value of $Y$ is indicated by the 
vertical line at which the respective branches end with $Q=0$, while the curve
with largest maximal $Q$ corresponds to the smallest value of $\kappa$ (and vice versa). 
Starting from $\omega_{max}$ with $Q=0$ and progressing along the
branch the solutions possess ergoregions from the dot onwards.}
\end{figure}

Since our solutions are rotating, we would expect that 
ergoregions appear. This would make it possible to amplify incident scalar waves
in a process very similar to that for rotating black holes. Since the conformal boundary of AdS provides a ``natural mirror'' on which
waves would bounce of and interact again with the interior of the space-time it has been suggested that in the latter case this could lead
to a ``black hole bomb'' \cite{Press:1972zz,Cardoso:2004nk}.
In asymptotically flat space-time it has been argued that the appearance of ergoregions leads
to a superradiant instability of boson stars \cite{Cardoso:2007az}. We believe that a very similar argument holds in
AdS, since it is the behaviour of the metric functions close to the boson star core that matters for this instability
to appear and not the asymptotic behaviour of the space-time.

The ergoregion of the rotating boson star in 5 dimensions is different from that appearing e.g. in the 4-dimensional Kerr space-time.
In the latter, the radius of the ergosphere corresponding to the outer surface of the ergoregion depends on the angular variable $\theta$.
This is related to the fact that in 4 dimensions only one plane orthogonal to the axis of rotation exists. In 5 dimensions
two such planes exist. In this paper we have chosen both angular momenta of the solution to be equal. This enhances the symmetry
and hence the radius of the ergosphere is independent of angular coordinates. The role that the two equal angular momenta play
is also seen when considering the form of the metric tensor, see (\ref{metric}). The $g_{tt}$ component of the metric is
independent of $\theta$ since the terms depending on $\theta$ in the two angular-tenporal parts add up to one such that
$g_{tt}=-b(r)+h(r)W^2(r)$. Hence, the condition for the ergoregion, which reads $g_{tt}\geq 0$ (the ergosphere corresponding
to the equality sign), is also independent of $\theta$. Note also that there is a crucial difference between the ergoregion
appearing for black holes and that for boson stars. The Kerr black hole always possesses an ergoregion. This is different
for boson stars. Rotating boson stars have an ergoregion only under specific conditions, which we will elaborate on
in the following.

Our numerical results indicate that ergoregions can appear for large values of the slope of the scalar field at the origin, $\phi'(0)$ as well
as sufficiently large values of $Y$.
This is shown for $\kappa=0.025$ in Fig.\ref{stability2}, where we give $M/Q$ as function of $J$ for two different
values of $Y$. While for $Y=0.0025$ we observe no ergoregion in the range of $J$ values plotted, an ergoregion appears on
the second branch of solutions for $Y=0.0050$. This is the solution marked by the cross at the upper left corner of Fig.\ref{stability2}.
In order to understand this better, we have also plotted the metric and scalar field functions for $\kappa=0.025$ and $Y=0.0050$ for
different values of $\phi'(0)$. In Fig.\ref{ergo_functions} we show the $g_{tt}$ component of the metric, the metric functions $f(r)$, $b(r)$, $h(r)$, $W(r)$ as
well as the scalar field functions for the solutions marked by the crosses in Fig.\ref{stability2}.
We find that for this choice of parameters and for $\phi'(0) \gtrsim 10$ the metric component $g_{tt}$ 
becomes larger than zero in a small region around the origin (see Fig.\ref{g_tt}).
In this ergoregion the
asymptotically time-like Killing vector becomes space-like.
Interestingly, this happens
for reasonably small angular momenta ($J=2100$) corresponding to the cross close to the top of Fig.\ref{stability2}. 
We can also see what happens to the individual metric functions. The minimum of $f(r)$ shifts to smaller values of $r$ and
at the same time has smaller values when increasing the value of $\phi'(0)$, while the metric function $b(r)$ has a decreasing value at the origin.
The maximum of the function $h(r)/r^2$ increases, the location of the maximum of $h(r)/r^2$ shifts inwards and the value of the function $W(r)$ at the origin 
decreases. Finally, it is very interesting to discuss the behaviour of the scalar field function.
For increasing $\phi'(0)$ the maximum of the scalar field function increases, while at the same time the maximum is located
at smaller values of $r$. Hence, we conclude that in order for an eregoregion to appear we need to have the scalar field
concentrated in a spherical shell close enough to the origin and at the same time have large enough maximal values of the scalar field
in this spherical shell.
This also explains why it is easier to obtain an ergoregion when increasing $Y$: choosing $Y$ larger essential means
decreasing the AdS radius in comparison to the physical length scale of the scalar field $\sim 1/m^2$ and hence
confining the scalar field in a smaller AdS ``box''.

Note that following our arguments about the stability, these are solutions on the second branch of solutions
which we would hence expect to be unstable evoking arguments from catastrophe theory. This seems to be a generic feature. 
Hence, in Fig.\ref{stability_details} we show for which values of the mass and charge the ergoregions appear. 
In Fig.\ref{ergo_detail1}
we show the region around $\omega=1$ in more detail by plotting $Q$ as function of $\bar{\omega}$ with 
$\bar{\omega}:=(\exp(\omega-1))^3$. 

In order to understand whether a boson star solution with an ergoregion could hence form at all requires a 
study of the full time-dependent problem modelling the formation of boson stars in a dynamical process. It can surely not be
excluded that unstable 
boson star solutions form. The question that remains to be answered is then whether these unstable boson stars
exist long enough to be interesting. To answer this question is, however, beyond the scope of this paper.

\section{Conclusions}
In this paper we have presented self-interacting boson stars with only one Killing vector field in aAdS. 
In contrast to previous studies of rotating boson stars in aAdS our solutions are composed of massive and self-interacting
scalar fields. We find that explicit solutions of the scalar field equation in a fixed AdS background exist
if the scalar field possesses mass only and no self-interaction. These are the generalisations of the oscillons
previously discussed for the massless case \cite{Bizon:2011gg,Jalmuzna:2011qw,Dias:2012tq}.
The self-interaction and/or back-reacting space-time render the equations of motion non-linear and
the solutions have to be constructed numerically. 
We have studied the solutions in dependence on the (dimensionless) gravitational
coupling and the AdS radius, which are both scaled to unity in \cite{Dias:2011at,Stotyn:2013yka,Henderson:2014dwa}. We construct fundamental as well
as radially excited rotating boson stars and discuss how the spectrum depends on the dimensionless parameters. 
We find that in the back-reacting case in general two branches of solutions exist out of which one connects to the AdS vacuum
such that the solutions are linearly stable. The solutions on the second branch, however, should be linearly unstable.
On this second branch of solutions we find that ergoregions appear for large enough central density of the boson star. This
likely signals the existence of a superradiant instability of the solutions.

While aAdS boson stars composed of a massless scalar field are non-linearly stable \cite{Buchel:2013uba} we are planing to study the non-linear
stability of out solutions in future work. wwe believe that the existence of a discrete spectrum of massive
oscillons will help to do so. \\
\\ 
{\bf Acknowledgments} BH and JR 
gratefully acknowledge support within the framework of the DFG Research
Training Group 1620 {\it Models of gravity}. The work of YB  was supported in part by an ARC contract n° AUWB-2010/15-UMONS-1.

\end{document}